\documentclass[aoas,MSNbibl,nameyear,rotating,dvips]{arximspdf}
\usepackage{stfloats}
\usepackage{graphicx}

\doi{10.1214/13-AOAS654} 
\volume{7}
\issue{3}
\pubyear{2013}
\firstpage{1249}
\lastpage{1285}

\makeatletter
\fnbelowfloat
\newcommand{\rrvert}{\vert}
\newcommand{\llvert}{\vert}
\def\cal{\mathcal}

\newcommand{\like}{{\cal L}}
\newcommand{\pval}{$p$-value}
\newcommand{\Eth}{E_{\mathrm{th}}}
\newcommand{\Dmax}{D_{\mathrm{max}}}
\newcommand{\Nsrc}{N_S}

\newcommand{\tdrxn}{\omega} 
\newcommand{\hdrxn}{\varpi} 
\newcommand{\expo}{\varepsilon} 
\newcommand{\tlabel}{\lambda} 
\newcommand{\Fvec}{\mathbf{F}} 
\newcommand{\Aperp}{A_\perp} 

\newcommand{\gta}{\gtrsim}
\newcommand{\lta}{\lesssim}

\makeatother

\begin{document}
\begin{frontmatter}

\title{Multilevel Bayesian framework for modeling the production,
propagation and detection of ultra-high energy cosmic rays\thanksref{T1}}
\thankstext{T1}{Supported by NSF Grants AST-0908439 and DMS-08-05975.}
\runtitle{Multilevel models for cosmic rays}

\begin{aug}
\author[a]{\fnms{Kunlaya} \snm{Soiaporn}\corref{}\ead[label=e1]{ks354@cornell.edu}},
\address[a]{K. Soiaporn\\
Department of Operations Research\\
\quad and Information Engineering\\
Cornell University\\
288 Rhodes Hall\\
Ithaca, New York 14853\\
USA\\
\printead{e1}}
\author[b]{\fnms{David} \snm{Chernoff}\ead[label=e3]{chernoff@astro.cornell.edu}},
\address[b]{D. Chernoff\\
Department of Astronomy\\
Cornell University\\
602 Space Sciences Building\\
Ithaca, New York 14853\\
USA\\
\printead{e3}\hspace*{22.5pt}}
\author[c]{\fnms{Thomas} \snm{Loredo}\ead[label=e2]{loredo@astro.cornell.edu}},
\address[c]{T. Loredo\\
Department of Astronomy\\
Cornell University\\
620 Space Sciences Building\\
Ithaca, New York 14853\\
USA\\
\printead{e2}}
\author[d]{\fnms{David} \snm{Ruppert}\ead[label=e4]{dr24@cornell.edu}}
\address[d]{D. Ruppert\\
Department of Operations Research\\
\quad and Information Engineering\\
Cornell University\\
1170 Comstock Hall\\
Ithaca, New York 14853\\
USA\\
\printead{e4}}
\and
\author[e]{\fnms{Ira} \snm{Wasserman}\ead[label=e5]{ira@astro.cornell.edu}}
\address[e]{I. Wasserman\\
Department of Astronomy\\
Cornell University\\
626 Space Sciences Building\\
Ithaca, New York 14853\\
USA\\
\printead{e5}}
\affiliation{Cornell University}
\runauthor{K. Soiaporn et al.}
\end{aug}

\received{\smonth{6} \syear{2012}}
\revised{\smonth{3} \syear{2013}}

%
\begin{abstract}
Ultra-high energy cosmic rays (UHECRs) are atomic nuclei with energies
over ten million times energies accessible to human-made particle
\mbox{accelerators}. Evidence suggests that they originate from relatively
nearby extragalactic sources, but the nature of the sources is unknown.
We develop a multilevel Bayesian framework for assessing association of
UHECRs and candidate source populations, and Markov chain Monte Carlo
algorithms for estimating model parameters and comparing models by
computing, via Chib's method, marginal likelihoods and Bayes factors.
We demonstrate the framework by analyzing measurements of 69 UHECRs
observed by the Pierre Auger Observatory (PAO) from 2004--2009, using a
volume-complete catalog of 17 local active galactic nuclei (AGN) out to
15~megaparsecs as candidate sources. An early portion of the data
(``period~1,'' with 14 events) was used by PAO to set an energy cut
maximizing the anisotropy in period~1; the 69 measurements include this
``tuned'' subset, and subsequent ``untuned'' events with energies above
the same cutoff. Also, measurement errors are approximately summarized.
These factors are problematic for independent analyses of PAO data.
Within the context of ``standard candle'' source models (i.e., with a
common isotropic emission rate), and considering only the 55 untuned
events, there is no significant evidence favoring association of UHECRs
with local AGN vs. an isotropic background. The highest-probability
associations are with the two nearest, adjacent AGN, Centaurus~A and NGC~4945.
If the
association model is adopted, the fraction of UHECRs that may be
associated is likely nonzero but is well below 50\%. Our framework
enables estimation of the angular scale for deflection of cosmic rays by
cosmic magnetic fields; relatively modest scales of $\approx\!3^\circ$ to
$30^\circ$ are favored. Models that assign a large fraction of UHECRs
to a single nearby source (e.g., Centaurus~A) are ruled out unless very
large deflection scales are specified a priori, and even then they are
disfavored. However, including the period~1 data alters the conclusions
significantly, and a simulation study supports the idea that the
period~1 data are anomalous, presumably due to the tuning. Accurate and
optimal analysis of future data will likely require more complete
disclosure of the data.
\end{abstract}

%
\begin{keyword}
\kwd{Multilevel modeling}
\kwd{hierarchical Bayes}
\kwd{astrostatistics}
\kwd{cosmic rays}
\kwd{active galactic nuclei}
\kwd{directional data}
\kwd{coincidence assessment}
\kwd{Bayes factors}
\kwd{Chib's method}
\kwd{Fisher distribution}
\end{keyword}

\end{frontmatter}

\section{Introduction}

Cosmic ray particles are naturally produced, positively charged atomic nuclei
arriving from outer space with velocities close to the speed of light.
The origin of cosmic rays is not well understood. The Lorentz force
experienced by a charged particle in a magnetic field alters its trajectory.
Simple estimates imply that cosmic rays with energy $E \lta10^{15}$~eV
have trajectories so strongly bent
by the Galactic magnetic field that they are largely trapped within the
Galaxy.\setcounter{footnote}{1}\footnote{An electron volt (eV) is the energy gained by an
electron accelerated through a 1 Volt potential; the upgraded Large Hadron
Collider will accelerate protons to energies $\sim\!7\times10^{12}$~eV. We
follow the standard astronomical convention of using ``Galaxy'' and
``Galactic'' (capitalized) to refer to the Milky Way galaxy.}
The acceleration sites and the source populations are not definitively known
but probably include supernovae, pulsars, stars with strong winds and
stellar-mass black holes. For recent reviews, see \citet{Cronin99} and
\citet{Hillas06}.
More mysterious, however, are the highest energy cosmic rays.

By 1991, large arrays of cosmic ray detectors had seen a few events with
energies $\sim\!100$~EeV (where EeV $=10^{18}$~eV). In the 1990s the Akeno
Giant Air Shower Array [AGASA; \citet{1992APh.....1...27C}] and the High
Resolution Fly's Eye [HiRes; \citet{2002NIMPA.482..457B}] were built to
target these ultra-high energy cosmic rays (UHECRs); each detected a few
dozen cosmic rays with $E>10$~EeV. For recent reviews, see
\citet{KO11-UHECRs,LS11-UHECRs,KW12-UHECRs-hist}.
Detectable UHECRs likely emanate from relatively nearby extragalactic
sources. On the one hand, their trajectories are only weakly deflected by
galactic magnetic fields so they are unconfined to the galaxy from which
they originate. On the other hand, they are unlikely to reach us from
distant (and thus isotropically distributed) cosmological sources.
Cosmic ray protons with energies above the Greisen--Zatsepin--Kuzmin (GZK)
scale of $\sim\!50$ to 100~EeV should scatter off of cosmic microwave
background photons, losing some of their energy to pion production with each
interaction [\citet{G66-GZK,ZK66-GZK}]; heavier nuclei can lose energy from
other interactions at similar energy scales.
Thus, the universe is not transparent to UHECRs;
they are not expected to travel more than $\sim\!100$ megaparsecs (Mpc;
a parsec is $\approx\!3.26$ light years)
before their energies fall below the GZK scale. Notably, over this distance
scale there is significant anisotropy in the distribution of matter that
should be reflected in the arrival directions of UHECRs. Astronomers hope
that continued study of the directions and energies of UHECRs will address
the fundamental questions of the field: What phenomenon accelerates
particles to such large energies? Which astronomical objects host the
accelerators? What sorts of nuclei end up being energized? In addition,
UHECRs probe galactic and intergalactic magnetic fields.

The flux of UHECRs is very small, approximately 1 per square kilometer per
century for energies $E\gta50$~EeV. Large detectors are needed to find
these elusive objects; the largest and most sensitive
detector to date is the Pierre Auger Observatory [PAO; \citet{PAO04-Proto}]
in Argentina. The
observatory uses air fluorescence telescopes and water
Cerenkov surface detectors to observe the air shower generated when a cosmic
ray interacts with nuclei in the upper atmosphere over the
observatory. The surface detectors (SDs) operate continuously, detecting
energetic subatomic particles produced in the air shower and reaching the
ground. The fluorescence detectors (FDs) image light from the air
shower and
supplement the surface detector data for events detected on clear, dark
nights.\footnote{The FD on-time is about 13\% [\citet{PAO10-GZK}],
but analysis
can reveal complications preventing use of the data---for example,
obscuration due to
light cloud cover or showers with significant development underground---so
fewer than 13\% of events have usable FD data. These few so-called
\emph{hybrid} events are important for calibrating energy measurements
and provide
information about cosmic ray composition vs. energy.}
PAO began taking data in 2004 during construction; by
June 2008 the PAO array comprised~$\approx\!1600$ SDs
covering $\approx\!3000$ km$^2$, surrounded by four
fluorescence telescope stations (with six telescopes in each station)
observing the atmosphere over the array.

By 31 August 2007, PAO had detected 81 UHECRs with $E > 40$~EeV
[see \citet{PAO07-Aniso}, hereafter PAO-07], finding clear evidence of
an energy cutoff resembling the predicted GZK cutoff, that is, a sharp
drop in
the energy spectrum above~$\approx\!100$~EeV and a discernable pile-up of
events at energies below that [\citet{PAO10-GZK}]. This supports the
idea that
the UHECRs originate in the nearby universe, although other
interpretations are possible.\footnote{The PAO data also indicate that the
composition of cosmic rays changes with energy, with protons dominant at
$E\approx\!1$~EeV but heavier nuclei becoming predominant for $E\gta10$~EeV
[\citet{PAO10-CRComposn,KU12-CRComposn}]. Astrophysically, it is
natural to presume that the maximum energy a cosmic accelerator can
impart to a nucleus of charge $Z$ grows with $Z$. Combined with the
PAO composition measurements, this has motivated models for which
the maximum energy for protons is $\sim\!1$~EeV, with the observed
cutoff above 50~EeV reflecting the maximum energy for heavy nuclei
[\citet{A+08-CRCompEmax,A+12-UHECRDis}].
In such models, there is no GZK suppression; the observed cutoff
reflects properties of the cosmic ray acceleration process.}

The PAO team searched for correlations between the cosmic ray arrival
directions and the directions to nearby active galactic nuclei (AGN)
[initial results were reported in PAO-07; further details and a catalog of
the events are in \citet{PAO08-AGN}, hereafter PAO-08]. AGN are unusually
bright cores of galaxies; there is strong (but indirect) evidence that they
contain rapidly mass-accreting supermassive black holes that eject some
material in energetic, jet-like outflows. AGN are theoretically favored
sites for producing UHECRs; electromagnetic observations indicate particles
are accelerated to high energies near AGN. The PAO team's analysis was
based on a significance test that counted the number of UHECRs with best-fit
directions within a critical angle, $\psi$, of an AGN in a catalog of local
AGN (more details about the catalog appear below); the number was compared
with what would be expected from an isotropic UHECR directional distribution
using a \pval. A simple sequential approach was adopted. The earliest half
of the data was used to tune three parameters defining the test
statistic by
minimizing the \pval. The parameters were as follows: $\psi$; a
maximum distance,
$\Dmax$, for possible hosts; and a minimum energy, $\Eth$, for UHECRs
considered to be associated with AGN. With these parameters tuned
($\Eth=56$~EeV, $\psi=3.1^\circ$, $\Dmax=75$ Mpc), the test was
applied to
the later half of the data; 13 UHECRs in that period had $E>\Eth$. The
resulting \pval\ of $1.7\times10^{-3}$ was taken as indicating the data
reject the hypothesis of isotropic arrival directions ``with at least a
99\%
confidence level.'' The PAO team was careful to note that this result did
not necessarily imply that UHECRs were associated with the cataloged AGN,
but rather that they were likely to be associated with some nearby
extragalactic population with similar anisotropy.

Along with these results, the PAO team published a catalog of energy and
direction estimates for the 27 UHECRs satisfying the $E>\Eth$ criterion,
including both the earliest 14 events used to define $\Eth$ and the 13
subsequent events used to obtain the reported \pval\ (the PAO data are
proprietary; measurements of the other 54 events used in the analysis were
not published). Their statistical result
spurred subsequent analyses of these early published PAO UHECR
arrival directions, adopting different methods and aiming to make more
specific claims about the hosts of the UHECRs. Roughly speaking, these
analyses found similarly suggestive evidence for anisotropy, but no
conclusive evidence for any specific association hypothesis.


In late 2010, the PAO team published a revised catalog, including new data
collected through 2009 [\citet{PAO10-AnisoUpdate}; hereafter PAO-10]. An
improved analysis pipeline revised the energies of earlier events downward
by 1~EeV; accordingly, the team adopted $\Eth= 55$~EeV on the new energy
scale. The new catalog includes measurements of 42 additional UHECRs (with
$E>\Eth$) detected from 1 September 2009 through 31 December 2010. A repeat
of the previous analysis (adding the new events but again excluding the
early tuning events) produced a larger \pval\ of $3\times10^{-3}$,
that is,
\emph{weaker} evidence against the isotropic hypothesis. The team performed
a number of other analyses (including considering new candidate host
populations). Despite the growth of the post-tuning sample size from 14 to
55, they found that the evidence for anisotropy weakened. Time-resolved
measures of anisotropy provided puzzling indications that later data
might have different directional properties than early data, although
the sample size is too small to demonstrate this conclusively.
Various investigators have performed other analyses aiming to detect
anisotropy in the distribution of detected UHECR directions, the
vast majority also adopting a hypothesis testing approach (seeking to
reject isotropy), but differing in choices of test statistic.
Most such tests require some accounting for tuning parameters, and many
do not explicitly account for measurement errors. See \citet{KK11-AGN-UHECR}
for a recent example with references to numerous previous frequentist analyses.

Here we describe a new framework for modeling UHECR data based on Bayesian
multilevel modeling of cosmic ray emission, propagation and detection.
A virtue of this approach is that physical and experimental
processes have explicit representations in the framework, facilitating
exploration of various scientific hypotheses and physical
interpretation of
the results.
This is in contrast to
hypothesis testing approaches where elements such as the choice of
test statistic or angular and energy
thresholds, only implicitly represent underlying physics, and potentially
conflate astrophysical and experimental effects (e.g., magnetic scattering
of trajectories and measurement errors in direction).
Our framework can
handle a priori uncertainty in model parameters via marginalization.
Marginalization also accounts for the uncertainty in such parameters via
weighted averaging, rather than fixing them at precise, tuned values. This
eliminates the need to tune energy, angle and distance scales with a subset
of the data that must then be excluded from a final analysis. Such parameters
are allowed to adapt to the data, but the ``Ockham's razor'' effect associated
with marginalization penalizes models for fine-tuned degrees of freedom,
thereby accounting for the adaptation.

Our approach builds on our earlier work on Bayesian assessment of
spatiotemporal coincidences in astronomy (see Section~\ref{sec3}). A recent
approximate Bayesian analysis of coincidences between UHECR and AGN
directions independently adopts some of the same ideas [\citet
{WMJ11-BayesUHECR}];
we discuss how our approach compares with this recent analysis in
the supplementary material [\citet{S+13-UHECR-Supp}].

In this paper we describe our general framework, computational algorithms
for its implementation and results from analyses based on a few
representative models. Our models are somewhat simplistic astrophysically,
although similar to models adopted in previous studies. We do not aim to
reach final conclusions about the sources of UHECRs; the focus here is on
developing new methodology and demonstrating the capabilities of the
approach in the context of simple models.

An important finding is that \emph{thorough and accurate independent analysis
of the PAO data likely requires more data than has so far been publicly
released} by the PAO collaboration. In particular, although our Bayesian
approach eliminates the need for tuning, in the absence of publicly available
``untuned'' data (i.e., measurements of lower-energy cosmic rays), we cannot
completely eliminate the effects of tuning from analyses of the
published data (Bayesian or otherwise). Additionally, a~Bayesian
analysis can
(and should) use event-by-event (i.e., heteroskedastic) measurement
uncertainties, but these are not publicly available. Finally, astrophysically
plausible conclusions about the sources of UHECRs will require models more
sophisticated than those we explore here (and those explored in other recent
studies).

\section{Description of cosmic ray and candidate host data}

\subsection{Cosmic ray data}
\label{sec:data}

The reported PAO measurements depend not only on the intrinsic particle
population but also on many experimental and algorithmic choices in the
detection and analysis chain, many of them associated with the need to
distinguish between events of interest and background events from uninteresting
but uncontrollable sources (e.g., natural radioactivity). UHECRs can
impinge on the observatory at any time, from any direction and with any
energy.
However, virtually no background sources
produce events with properties mimicking those of very high energy
cosmic rays
arriving from directions well above the horizon.
Cosmic rays with $E>3$~EeV arriving from any direction lying within a large
window on the sky create air showers detected with nearly 100\% efficiency
(no false positives, no false dismissals). The SDs and FDs measure
the spatiotemporal development of the air shower which allows
the energy and arrival direction to be measured. The uncertainties depend
upon how many counters of each type are triggered plus the systematic and
statistical uncertainties implicit in modeling the development of the
air shower. The PAO team reports energy and
arrival direction estimates for each cosmic ray falling within the geometric
bounds of its zone of secure detection.\footnote{The directional criterion
adopted for the PAO catalogs is that an event is reported if its best-fit
arrival direction is within $60^\circ$ of the observatory's zenith, the
local normal to Earth's surface at the time of the event.}

We consider the $N_C = 69$ UHECRs with energies $E \geq\Eth= 55$~EeV
cataloged in PAO-10, which reports measurements of all UHECRs seen by PAO
through 31 December 2009 with $E \geq\Eth$, based on analysis of the
surface detector data only. Although our framework does not tune an event
selection criterion, for interpreting the results it is important to
remember that the $\Eth=55$~EeV threshold value was set to maximize a
signature of anisotropy in an early subset of the data.
The tuning data included the 14 earliest
reported events, detected from 1 January 2004 to 26 May 2006 (inclusive;
period 1), as well as numerous unreported events with $E<\Eth$. The first
published catalog in PAO-08 included 13 subsequent UHECRs observed
through 31
August 2007 (period 2). The PAO-10 catalog includes 42 additional UHECRs
observed through 31 December 2009 (period 3). Table~1 in PAO-10 provides
information about the three periods, including the sky exposure for each
period, which is not simply proportional to duration (the observatory
grew in
size considerably through 2008). Data for cosmic rays with $E < \Eth$
are not
publicly available.\footnote{The PAO web site hosts public data for
1\% of
lower-energy cosmic rays, but the sample is not statistically characterized
and UHECRs are not included.}

The direction estimate for a particular cosmic ray is the result of a
complicated analysis of time series data from the array of PAO surface
detectors.\footnote{The SD data may be supplemented by data from the
fluorescence detectors for hybrid events observed under favorable
conditions, but there are very few such events at ultra-high energies,
and PAO-10 reports analysis of SD data only.}
Roughly speaking, the direction is inferred by triangulation.
The analysis produces a likelihood function for the cosmic ray arrival
direction, $\omega$ (a unit vector on the celestial sphere). The
shapes of
the likelihood contours are not simple, but they are roughly azimuthally
symmetric about the best-fit direction. The PAO-10 catalog summarizes the
likelihood function with a best-fit direction and a typical directional
uncertainty of $\approx\!0.9^\circ$ corresponding to the angular radius
of an
azimuthally symmetric 68.3\% confidence region. We use these summaries to
approximate the likelihood functions with a Fisher distribution with
mode at
the best-fit direction for each cosmic ray, and with concentration parameter
$\kappa_c = 9323$, corresponding to a 68.3\% confidence region with an
angular radius of 0.9$^\circ$. Let $d_i$ denote the data associated with
cosmic ray $i$, and $\omega_i$ denote its actual arrival direction (an
unknown parameter). The
likelihood function for the direction is
%
%
\begin{equation}
\ell(\omega_i):= P(d_i|\omega_i) \approx
\frac{\kappa_c}{4\pi\sinh(\kappa_c)} \exp(\kappa_c n_i\cdot
\omega_i), \label{ell-def}
\end{equation}
where $n_i$ denotes the best-fit direction for cosmic ray $i$ (a function
of the observed data), and we have
scaled the likelihood function so its integral over $\omega_i$ is unity,
merely as a convenient convention.
\citet{B+PAO09-DrxnUncert} provide more information about the PAO direction
measurement capability. Note that the expected angular scale of magnetic
deflection is larger than the PAO directional uncertainties, significantly
so if UHECRs are heavy nuclei (see Section~\ref{sec:dflxn}).

Similarly, the analysis pipeline produces energy estimates for each event.
These estimates have significant random and systematic uncertainties
[Abraham et al. (\citeyear{PAO08-GZK,PAO10-GZK})].
The models we study here do not make use of the reported energies and are
unaffected by these uncertainties. But our framework readily
generalizes to
account for energy dependence. In principle, it is straightforward to account
for the random uncertainties, but a consistent treatment requires data for
events below any imposed threshold: the true energies of events with best-fit
energies below threshold could be above threshold (and vice versa for those
with best-fit energies above threshold); accounting for this requires
data to
energies below astrophysically important thresholds. The systematic
uncertainties become important for joint analyses of PAO data with data from
other experiments, and for linking results of spectral analyses to particle
physics theory.

\subsection{Candidate source catalog}

As candidate sources for the PAO UHECRs, the analysis reported in
PAO-07 and
PAO-08, and several subsequent analyses, considered 694
AGN within $\approx\!75$ Mpc from the 12th catalog
assembled by V\'eron-Cetty and V\'eron [VCV; \citet{VCV-12thAGNCat}].
This catalog includes data on all AGN and quasars (AGN with star-like
images) with published spectroscopic redshifts; it includes observations
from numerous investigators using diverse equipment and AGN selection
methods, and does not represent a statistically well-characterized
sample of
AGN.\footnote{VCV say of the catalog, ``This catalogue should not be used
for any statistical analysis as it is not complete in any sense, except that
it is, we hope, a complete survey of the literature.''}
Subsequent analyses in PAO-10, and a few other analyses, used more recent
catalogs of active galaxies or normal galaxies, including flux-limited
catalogs (i.e., well-characterized catalogs that contain all bright sources
within a specified volume, but dimmer sources only in progressively smaller
volumes).


For the representative analyses reported here, we consider the 17 AGN
cataloged by \citeauthor{2010MNRAS.406..597G} [(\citeyear{2010MNRAS.406..597G});
hereafter G10]
as candidate sources. This is a well-characterized \emph{volume}-limited
sample; it includes all infrared-bright AGN within 15 Mpc.
For each AGN in the calalog, we take its position on the sky, $\varpi_k$
($k=1$ to $\Nsrc$), and its distance, $D_k$, to be known
precisely.\footnote{Galaxy directions have negligible uncertainties compared
to cosmic ray directions. Three of the AGN have distances measured
using the
Cepheid variable period-luminosity relation, and four others have distances
inferred from one or more of the following distance indicators:
Tully--Fisher, surface brightness fluctuations, type Ia Supernovae, and
fundamental plane. The remaining AGN have distances inferred from redshifts
using a local dynamical model. The errors likely range from
one to a few Mpc, small enough to be inconsequential for our analyses.}
Notably, this catalog includes Centaurus A (Cen A), the nearest AGN
($D\approx4.0$ Mpc), an unusually active and morphologically peculiar AGN.
Theorists have hypothesized Cen A to be a source of many or even most UHECRs
if UHECRs are heavy nuclei, which would be deflected through large angles;
see \citet{B+09-CenA,GBdS10-CenA,BdS12-CenA}. The small size of this catalog
facilitates thorough exploration of our methodology: Markov chain Monte
Carlo algorithms can be validated against more straightforward algorithms
that could not be deployed on large catalogs, and simulation studies are
feasible that would be too computationally expensive with large catalogs.
Also, for simple ``standard candle'' models (adopted here and in other
studies) that assign all sources the same cosmic ray intensity, little is
gained by considering large catalogs, because assigning detectable cosmic
ray intensities to distant sources would imply cosmic ray fluxes from nearby
sources too large to be compatible with the data.

We also include an isotropic background component as a ``zeroth'' source.
This allows a model to assign some UHECRs to sources not included in
the AGN
catalog (either galaxies not cataloged or other, unobserved sources). In
addition, we consider an isotropic source distribution for \emph{all} cosmic
rays (i.e., a model with only the zeroth source) as a ``null'' model for
comparison with models that associate some cosmic rays with AGN or other
discrete sources. An isotropic distribution is convenient for calculations
and has been adopted as a null hypothesis in several previous studies.
Historically, before PAO's convincing observation of a GZK-like cutoff in
the UHECR energy spectrum, the isotropic distribution was meant to represent
a distant cosmological origin for UHECRs. Accepting the null would indicate
that the GZK prediction was incorrect and that changes in fundamental
physics would be required to explain UHECRs. In light of PAO's compelling
observation of a GZK-like cutoff (with its implied $\sim\!100$ Mpc distance
scale), interpreting an isotropic null or background component is
problematical if there are many light nuclei among the UHECRs. We adopt it
here both for convenience and due to precedent. We discuss this further
below.

\begin{figure}

\includegraphics{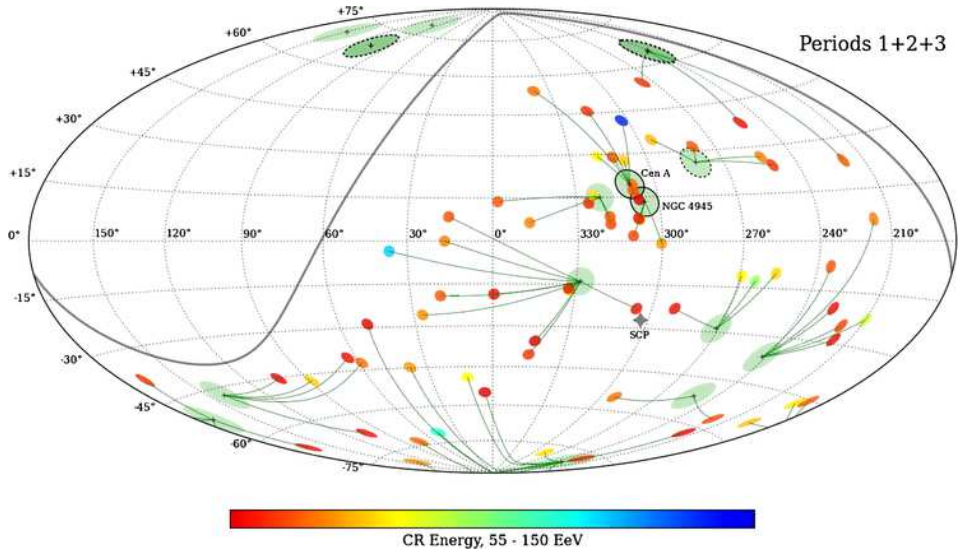}

\caption{Sky map showing directions to 69 UHECRs detected by PAO and
to 17 nearby AGN from the catalog of Goulding et al. Directions are
shown in an equal-area Hammer--Aitoff projection in Galactic coordinates.
Thick gray line indicates the boundary of the PAO field of view.
Small tissots show UHECR directions; tissot radius is $2^\circ$ corresponding
to $\approx\!2$ standard deviation errors; tissot color indicates energy.
Large green tissots indicate AGN directions; tissot radius is $5^\circ$.
Thin curves are geodesics connecting each UHECR to its nearest AGN.}
\label{fig:skymap}
\end{figure}

\subsection{Sky map}

Figure~\ref{fig:skymap} shows a sky map displaying the
directions to both the UHECRs seen by PAO and the AGN in the G10 catalog.
The directions are shown in an equal-area Hammer--Aitoff projection in
Galactic coordinates; the Galactic plane is the equator (Galactic latitude
$b=0^\circ$), and the vertically-oriented grid lines are meridians of
constant Galactic longitude, $l$. The star indicates the south celestial
pole (SCP), the direction directly above Earth's south pole (effects like
precession and nutation of the Earth's axis are negligible for this
application and we ignore them in this description). The thick gray line
bounds the PAO field of view. The UHECR and AGN directions are
displayed as
``tissots,'' projections of circular patches centered on the reported
directions. The small tissots show the UHECR directions; the tissot
size is
$2^\circ$, corresponding to $\approx\!2$ standard deviation errors, and the
tissot color indicates energy. The large green tissots indicate AGN
directions; the tissot size is $5^\circ$, corresponding to a plausible scale
for magnetic deflection of UHE protons in the Galactic magnetic
field.\footnote{If UHECRs are comprised of heavier, more positively charged
nuclei, they could suffer much larger deflections; see Section~\ref{sec:dflxn}.}
The tissots are rendered with transparency; the two darker tissots near the
Galactic north pole indicate pairs of AGN with nearly coincident directions.
Two of the AGN tissots are outlined in solid black; these correspond to the
two nearest AGN, Centaurus A (Cen A, also known as NGC 5128) and NGC 4945,
neighboring AGN at distances of 4.0 and 3.9 Mpc (as reported in G10). Five
others are outlined in dashed black; these have distances ranging from 6.6
to 10.0 Mpc (the two pairs of nearly coincident AGN are among these).
The remaining 10 AGN have distances from 11.5 to 15.0 Mpc.
Four of the AGN are outside the PAO field of view, but depending on the
scale of magnetic deflection, they could be sources of observable cosmic
rays.

Figure~\ref{fig:skymap} shows the measured directions for the 69 UHECRs.
The thin curves (teal) show
geodesics connecting each UHECR to its nearest AGN.
There is a noticeable concentration of cosmic ray directions near the
directions of Cen A and NGC 4945; a few other AGN also have conspicuously
close cosmic rays. We have also examined similar maps for the subsets
of the UHECRs in the three periods.
The concentration in the vicinity of the two closest AGN
is also evident in the maps for periods 1 and 2.
Curiously, except for a single UHECR about $6^\circ$
from NGC 4945, no such concentration is evident in the map for period 3,
despite it having about three times the number of UHECRs found in earlier
periods. This is a presage of results from our quantitative analysis that
suggest the data may not be consistent with simple models for the
cosmic ray
directions, with or without AGN associations.

\subsection{PAO exposure}
\label{sec:expo}

PAO is not equally sensitive to cosmic rays coming from all directions.
Quantitative assessment of evidence for associations or other
anisotropy must
account for the observatory's direction-dependent exposure.

Let $F$ be the cosmic ray flux at Earth from a source at a given direction,
$\tdrxn$, that is, the expected number of cosmic rays per unit time
per unit
area normal to $\tdrxn$. Then the expected number of rays detected in a
short time interval $dt$ is $F \Aperp(t,\tdrxn) \,dt$, where $\Aperp
(t,\tdrxn)$ is
the projected area of the observatory toward $\tdrxn$ at time $t$. The total
expected number of cosmic rays is given by integrating over $t$; it can be
written as $F\expo(\tdrxn)$, with the \emph{exposure map} $\expo
(\tdrxn)$
defined by
%
%
\begin{equation}
\expo(\tdrxn):= \int_T \Aperp(\tdrxn,t) \,dt;
\label{expo-def-main}
\end{equation}
the integral is over the time intervals when the observatory was operating,
denoted collectively by $T$.
The supplementary material [\citet{S+13-UHECR-Supp}] describes calculation of
$\expo(\tdrxn)$; the thick gray curves shown in the sky maps mark the
boundary of the region of nonzero exposure.

\section{Modeling the cosmic ray data}\label{sec3}


The basic statistical problem is to quantify evidence for associating some
number (possibly zero) of cosmic rays with each member of a candidate
source population. The key observable is the cosmic ray direction; a
set of
rays with directions near a putative host comprises a multiplet potentially
associated with that host. This gives the problem the flavor of model-based
clustering (of points on the celestial sphere rather than in a
Euclidean space),
but with some novel features:
\begin{itemize}
\item The model must account for measurement error in cosmic ray properties.

\item Observatories provide an incomplete and distorted sample of
cosmic rays,
so the model must account for random truncation and nonuniform thinning.

\item The most realistic astrophysical models imply a joint
distribution for
the properties of the cosmic rays assigned to a particular source that is
exchangeable rather than a product of independent distributions (as is the
case in standard clustering).

\item The number of cosmic rays is informative about the intensity
scale of
the cosmic ray sources so the binomial point process model underlying
standard generative clustering approaches is not appropriate.
\end{itemize}


To account for these and other complexities, we model the data using a
hierarchical Bayesian framework with four levels:
\begin{longlist}[1.]
\item[1.]\emph{Source properties}: At the top level we specify the properties
of the sources of cosmic rays. This may include the choice of a candidate
source population of identified objects (e.g., a particular galaxy
population) and/or specification of the properties of a population of
unidentified sources. For a given candidate source population, we must
specify source directions and cosmic ray intensities. The simplest case is
a standard candle model, with each source having the same cosmic ray
intensity. More generally, we may specify a (nondegenerate) distribution
of source intensities; this corresponds to specifying a ``luminosity
function'' in other astronomical contexts. For a population of unidentified
sources, we must specify a directional distribution (isotropic in the
simplest case) as well as an intensity distribution.
\item[2.]\emph{Cosmic ray production}: We model the production of cosmic rays
from each source with a marked Poisson point process model for latent cosmic
ray properties. The incident cosmic ray arrival times have a homogeneous
intensity measure in time, and the marks include the cosmic ray energies,
latent categorical labels identifying the source of each ray, and
possibly labels identifying the nuclear species of each ray (for models
with compositional diversity).
\item[3.]\emph{Cosmic ray propagation}: As cosmic rays propagate from
their sources, their directions may be altered by interaction with
cosmic magnetic fields, and their energies may be altered by interaction
with cosmic background radiation. We model magnetic deflection of
the rays by introducing latent variables specifying the source and
arrival directions, and parametric directional distributions describing
the relationships between these directions.
Here we adopt a simple phenomenological model with a single parameter
specifying a typical scattering scale between the source and arrival
directions. As the data become more abundant and detailed, the framework
can accommodate more complex models, for example, with parameters explicitly
describing cosmic magnetic fields and cosmic ray composition.
Interactions of the most energetic cosmic rays with cosmic background
photons can reduce the cosmic ray energy. The effect is significant
for rays with $E\gta100$~EeV traveling over distances $\gta100$ Mpc.
For the nearby sources in the G10 catalog, the effect is negligible
and we ignore it here, but we briefly discuss how it may be handled
via latent energy variables below.
\item[4.]\emph{Detection and measurement}: Last, we model detection and
measurement, accounting for truncation and thinning of the incident
cosmic ray flux and measurement errors for directions and energies.
\end{longlist}

\begin{figure}

\includegraphics{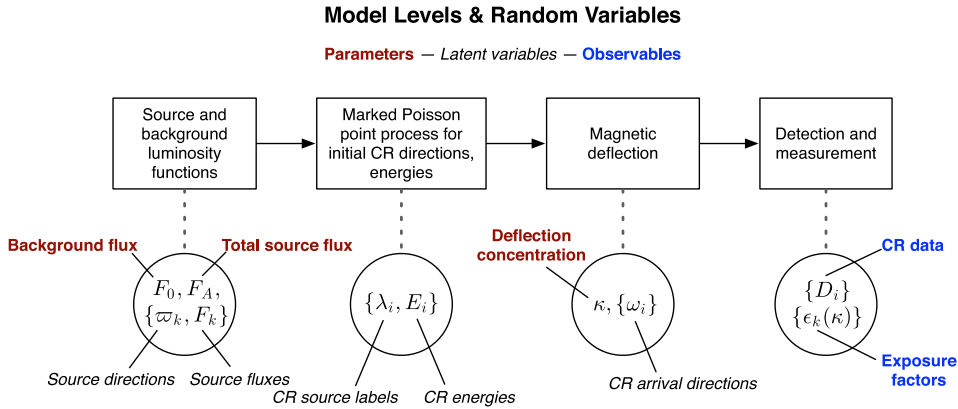}

\caption{Schematic depiction of the levels in our cosmic ray association
models, identifying random variables appearing in each level, including
parameters of interest (bold red labels), latent variables representing
cosmic ray
properties that are not directly observable (slant type labels) and
observables (bold blue labels).}
\label{fig:levels}
\end{figure}

Figure~\ref{fig:levels} schematically depicts the structure of our
framework, including identification of the various random variables
appearing in the calculations described below. The variables will be
defined as they appear in the detailed development below; the figure serves
as visual reference to the notation. The figure is not a graphical model
per se. Rather, our models specify probability distributions over a space
of graphs, each graph corresponding to a possible set of associations
of the
cosmic rays with particular sources. This framework builds directly on an
earlier multilevel Bayesian model we developed to assess evidence that some
sources of gamma-ray bursts repeat [\citet{LLW96}]; this model, too,
worked in
terms of probability distributions over candidate assignments. See
\citet{Loredo13-Coinc} for a broad discussion of Bayesian methods for
assessing spatiotemporal coincidences in astronomical data.

Our framework is designed to enable investigators to:
(1) Ascertain which cosmic rays (if any) may be associated with specific
sources with high probability; (2)~Estimate luminosity function parameters
for populations of astrophysical sources; (3) Estimate the proportion
of all
detected cosmic rays generated by each population; (4) Estimate parameters
describing the composition-dependent effects of cosmic magnetic fields;
(5) Investigate whether cosmic rays from a single source are deflected
independently or share part of their deflection history (resulting in
correlated deflections). Task (5) is not attempted here but will be
investigated in the future.

\subsection{Cosmic ray source properties}


We do not anticipate the UHECR flux passing through a volume element at the
Earth to vary in time over accessible time scales,\vadjust{\goodbreak} so we model the
arrival rate into a small volume of space from any particular direction
as a
homogeneous Poisson point process in time. Let $F_k$ denote the UHECR flux
from source $k$. $F_k$ is the expected number of UHECRs per unit time from
source $k$ that would enter a fully exposed spherical detector of unit
cross-sectional area. A cosmic ray source model must specify the directions
and fluxes of candidate sources. In our framework, a candidate source
catalog specifies source directions for a fixed number of potential sources,
$N_A$ ($N_A = 17$ for the G10 AGN catalog). In addition, we presume some
cosmic rays may come from uncatalogued sources, so we introduce a background
component, labeled by $k=0$, considered to be a population of isotropically
distributed ``background'' sources. We presume the background sources to
be numerous and to each have relatively low cosmic ray fluxes, so that at
most a single cosmic ray should be detected from any given background source
(i.e., we do not consider clustering of cosmic rays assigned to the
background). In this limit, the background component may be described
by a single parameter, $F_0$, denoting the total flux from the entire
background population.


A model must specify a distribution for $\{F_k\} = \{F_0, \Fvec\}$; in
astronomical jargon, this corresponds to specifying a ``luminosity function''
for the background and source populations. As a simple starting point, we
treat $F_0$ as a free parameter and adopt a ``standard candle'' model
specifying the $N_A$ candidate host fluxes, $\Fvec$, via a single parameter
as follows. We assume all sources emit isotropically with the same
intensity, $I$ (number of cosmic rays per unit time), so the flux from a
source (i.e., $F_k$ for $k>0$) can be written as $F_k = I/D_k^2$ (the
inverse-square law), with $D_k$ the (known) distance to source $k$ (there
could also be distance- and energy-dependent attenuation due to cosmic
ray--photon interactions, but the sources we consider here are close enough
that such attenuation should be negligible). The total flux from the
sources is $F_A = \sum_{k>0} F_k$, and we adopt $F_A$ as the source
intensity parameter rather than $I$. Thus, $F_k = w_k F_A$, with the weights
$w_k$ given by
%
%
\begin{equation}
w_k = \frac{1/D_k^2}{\sum_{j=1}^{N_A} 1/D_j^2} \label{wt-def}
\end{equation}
for $k=1$ to $N_A$.

\subsection{Top-level prior specification}\label{sec3.2}

We must specify a prior distribution for $F_0$ and $F_A$.
Earlier observations constrained the total UHECR flux. In our association
model, the total flux is $F_T = F_0 + F_A$. For the null model, there is
only one top-level parameter, the total flux from an isotropic distribution
of source directions. So we adopt $F_T$ as a top-level parameter,
common to all models. For association models, this motivates an alternative
parameterization that switches from $(F_0, F_A)$ to $(F_T, f)$, where
$f =
F_A/(F_0 + F_A)$ is the fraction of the total flux attributed to the
candidate host population.
In this parameterization, we can specify a common total flux prior
for all models. This is astrophysically sensible since we have
results from prior experiments to set a scale for the total flux.
It is also statistically desirable; Bayes factors tend to be robust
to specification of priors for parameters common to models being compared.

We adopt independent priors for the total flux and the associated fraction.
If their prior densities are $g(F_T)$ and $h(f)$, then the implied
joint prior density for $(F_0,F_A)$ is
%
%
\begin{equation}
\pi(F_0, F_A) = \frac{g(F_0+F_A) h ({F_A}/{(F_0+F_A)} )}{F_0+F_A}, \label{FF-prior}
\end{equation}
where the denominator is from the Jacobian of the transformation between
parameterizations. In general, an independent prior for $F_T$ and $f$
corresponds to a dependent prior for $F_0$ and $F_A$.

For the calculations below, we adopt an exponential prior with scale
$s$ for
$F_T$, and a beta prior for $f$ with shape parameters $(a,b)$, so
%
%
\begin{equation}
g(F_T) = \frac{1}{s}e^{-F_T/s}\quad \mbox{and}\quad h(f) =
\frac{1}{B(a,b)} f^{a-1}(1-f)^{b-1}, \label{exp-beta}
\end{equation}
where $B(a,b)$ is the beta function. We set the hyperparameters $(s,a,b)$
as follows.

We take $s = 0.01\times4\pi$ km$^{-1}$ yr$^{-1}$ for all models. This scale
is compatible with flux estimates from AGASA and HiRes. The likelihood
functions for $F_T$ from those experiments are formally different from
exponentials (they are more concentrated away from zero), but since this
prior is common to all models, and since the PAO data are very informative
about the total flux, our results are very robust to its detailed
specification.

For the beta prior for $f$, our default choice is $a=b=1$, which corresponds
to a uniform prior on $[0,1]$. We also repeat some computations using $b=5$
to investigate the sensitivity of Bayes factors to this prior. This case
skews the prior downward, increasing the probability that $f$ is close
to 0.

\subsection{Cosmic ray mark distributions}

Given the fluxes, we model cosmic ray arrival times with a
superposition of
homogeneous Poisson point processes from each component. Besides its
arrival time, each cosmic ray has a label associated with it, identifying
its source component. Let $\tlabel$ be an integer-valued latent label
for a
UHECR, specifying its source ($\tlabel= 0$ for the background or $k
\ge\!1$
for AGN $k$). Since a superposition of Poisson processes is a Poisson
process, we may consider the arrival times for the UHECRs arriving at Earth
to come from a total event rate process and the labels to come from a
categorical mark distribution with probability mass function
%
%
\begin{equation}
P(\lambda=k|F_0,\Fvec) = \frac{F_k}{\sum_{j=0}^{N_A} F_j}. \label{label-pmf}
\end{equation}
In the absence of magnetic deflection, the labels could be replaced by
source directions (with background source directions assigned
isotropically), and the process could be considered to be Poisson in time
with a directional mark distribution. But magnetic deflection requires a
more complex setup.

Our full framework also assigns energies as marks for each cosmic ray,
drawn from a distribution describing the emitted cosmic ray spectrum.
This potentially enables joint inference of directional and spectral
properties of cosmic ray sources.
The shape of the emitted spectrum reflects the physical processes
that produce UHECRs; introducing a parameterized emission spectrum
can allow the analysis to directly constrain production processes.
In addition, cosmic ray energies may be changed by interactions with
cosmic background photons during propagation, altering the spectrum.
When such effects are important, the measured energies provide indirect
information about the spatial distribution of cosmic ray sources. We
discuss this further in the supplementary material [\citet{S+13-UHECR-Supp}].
In the example analysis presented below, the candidate sources are
nearby, at distances~$\le\!15$ Mpc where propagation effects are negligible.
In addition, as explained in the supplementary material, the shape of the observed
spectrum at high energies can be intimately tied to its shape at
low energies (particularly for the isotropic component, which likely is
associated with distant sources). But PAO currently reports
measurements only for events with energies $\ge\!55$~EeV; the absence of
lower-energy data significantly compromises the ability to account for
propagation effects on the cosmic ray spectrum. For these reasons, in
the analysis presented here we ignore the energy mark distribution.
Analyses considering more distant candidate sources will have to
address these issues, along the lines described in the supplementary material.


\subsection{Propagation---magnetic deflection}
\label{sec:dflxn}

After leaving a source, UHECRs will have their paths deflected as they
traverse galactic and intergalactic magnetic fields. The Galactic field is
partially measured and is known to have both a turbulent component (varying
over length scales below $\sim\!1$ kpc) and a regular component (coherent
over kpc scales and largely associated with spiral arms), with typical field
strengths $\sim\!1\ \mu$G. The magnetic fields of other galaxies are at
best crudely measured and believed to be similar to the Galactic field. The
much smaller fields in intergalactic space are only weakly constrained (in
fact, cosmic rays might provide useful additional constraints); the typical
field strength is probably not larger than~$\sim\!10^{-9}$ G except within
galaxy clusters.

A number of investigators have modeled cosmic ray
propagation in the Galaxy, or in intergalactic space, using physical models
based on existing field measurements [recent examples include
\citet{HRM02-Lens,HMR02,D+05-CRDflxn,NM10-CRDflxn,AKP10-CRDflxn,J+10-CRSources};
see \citet{Sigl12} for an overview]. Roughly speaking, there are two regimes
of deflection behavior, described here in the small-deflection limit
[\citet{HMR02}]. As a cosmic ray with energy $E$ and atomic number $Z$
traverses
a distance $L$ spanning a regular magnetic (vector) field $\mathbf
{B}$, it is
deflected by an angle
%
%
\begin{equation}
\delta\approx 6.4^\circ Z \biggl(\frac{E}{50\ \mathrm{EeV}}
\biggr)^{-1} \biggl\llvert \int_L
\frac{d\mathbf{s}}{3\ \mathrm{kpc}} \times \frac{\mathbf{B}}{2\ \mu\mathrm{G}} \biggr\rrvert, \label{dflxn-reg}
\end{equation}
where $\mathbf{s}$ (a vector) is an element of displacement along the
trajectory; the field and length scales are typical for the Galaxy. If
instead it traverses a region with a turbulent structure, with the field
coherence length $\ell\ll L$, then the deflection will be stochastic; its
probability distribution has zero mean and root-mean-square (RMS) angular
scale
%
%
\begin{eqnarray}
\label{dflxn-turb} \delta_\mathrm{rms} &\approx& 1.2^\circ Z \biggl(
\frac{E}{50\ \mathrm{EeV}} \biggr)^{-1} \biggl(\frac{B_\mathrm{rms}}{4\ \mu\mathrm{G}} \biggr) \biggl(
\frac{L}{3\ \mathrm{kpc}} \biggr)^{1/2} \biggl(\frac{\ell}{50\ \mathrm{pc}}
\biggr)^{1/2}
\nonumber
\\[-8pt]
\\[-8pt]
\nonumber
&\approx& 2.3^\circ Z \biggl(\frac{E}{50\ \mathrm{EeV}} \biggr)^{-1}
\biggl(\frac{B_\mathrm{rms}}{1\ \mathrm{nG}} \biggr) \biggl(\frac{L}{10\ \mathrm{Mpc}} \biggr)^{1/2}
\biggl(\frac{\ell}{1\ \mathrm{Mpc}} \biggr)^{1/2},
\nonumber
\end{eqnarray}
where $B_\mathrm{rms}$ is the RMS field strength along the path, and quantities
are scaled to typical galactic and intergalactic scales on the first and
second lines, respectively.

For a detected cosmic ray, the energy is measured fairly accurately, but
other quantities appearing in the deflection formulae may be largely
unknown. As noted above, there is significant uncertainty in the magnitudes
of cosmic magnetic fields, particularly for turbulent structures. Turbulent
length scales are poorly known. Finally, the composition (distribution of
atomic numbers) of UHECRs is not known. Low energy cosmic rays are
known to
be mainly protons and light nuclei, but the proportion of heavy nuclei (with
$Z$ up to 26, corresponding to iron nuclei, the most massive stable nuclei)
increases with energy up to about $10^{15}$~eV. At higher energies,
inferring the cosmic ray composition is very challenging, requiring both
detailed measurement of air shower properties and theoretical modeling
of the $Z$ dependence of hadronic interactions at energies far beyond
those probed by accelerators. Measurements and modeling from HiRes
indicate light nuclei are predominant again at $\approx\!1$~EeV and remain
so at least to $\approx\!40$~EeV [\citet{HiRes10-Final-arxiv}].
In contrast, recent PAO measurements indicate a transition from light
to heavy nuclei over the range $\approx3$--30~EeV
[\citet{PAO10-CRComposn,PAO12-Composition}].
(The discrepancy is not yet explained.)
For heavy nuclei, the deflection scales in both the regular and
turbulent deflection regimes can be large, $\sim\!1$ rad. Some
investigators have suggested that many or most UHECRs may be heavy
nuclei originating from the nearest AGN, Cen A, so strongly deflected
that they come from directions across the whole southern sky [e.g.,
\citet{B+09-CenA,GBdS10-CenA,BdS12-CenA}].



In light of these uncertainties and the relative sparsity of UHECRs,
we use simple phenomenological models for magnetic deflection.
In the simplest ``buckshot'' model, each cosmic ray from a particular source
experiences a deflection that is conditionally independent of the deflection
of other rays from that source, given a parameter, $\kappa$,
describing the
distribution of deflections. We have also devised a more
complex ``radiant'' model that allows cosmic rays assigned to the same
source to have correlated deflections, with the correlation representing
a partially shared deflection history. For the analyses reported here, we
use the buckshot model; we describe the radiant model further in
Section~\ref{sec:summary}.

The buckshot deflection model adopts a Fisher distribution for the
deflection angles. The model has a single parameter, $\kappa$, the
concentration parameter for the Fisher distribution. The probability
density for observing a cosmic ray from direction $\tdrxn$ if it is assigned
to source $k$ with direction $\hdrxn_k$ is then
%
%
\begin{equation}
\rho_k(\tdrxn|\kappa) = \frac{\kappa}{4\pi\sinh(\kappa)}\exp(\kappa\tdrxn\cdot
\hdrxn_k). \label{rho-def}
\end{equation}
With this deflection distribution, when a cosmic ray is generated from an
isotropic background population, its deflected direction still has an
isotropic distribution. Accordingly,
%
%
\begin{equation}
\rho_0(\tdrxn|\kappa) = \frac{1}{4\pi}. \label{rho-iso}
\end{equation}

The $\kappa$ parameter is convenient for computation, but an angular scale
is more convenient for interpretation. The contour of the Fisher density
bounding a region containing probability $P$ is azimuthally symmetric
with angular radius $\theta_P$ satisfying
%
%
\begin{equation}
\int_{\Omega}\,d \tdrxn\, \rho_k(\tdrxn|\kappa)=
\frac{1-e^{-\kappa[1-\cos(\theta_P)]}}{1-e^{-2\kappa}}=P, \label{kappa-theta}
\end{equation}
where $\Omega$ denotes the cone of solid angle subtended by the
contour. In
plots showing $\kappa$-dependent results, we frequently provide an angular
scale axis, using (\ref{kappa-theta}) with $P=0.683$, in analogy to the
``$1\sigma$'' region of a normal distribution.\footnote{In the
$\kappa\gg1$
limit, the Fisher density becomes an uncorrelated bivariate normal with
respect to locally cartesian arc length coordinates about the mode on the
unit sphere.
The standard deviation in each of the coordinate directions is
$\sigma\approx1/\kappa^{1/2} \approx57.3^\circ/\kappa^{1/2}$ in
this limit.
The radius containing 68.3\% probability, $\theta_P$,
satisfies equation (\ref{kappa-theta}) with $P = 0.683$; for $\kappa
\gg
1$ this implies $\theta_P^2 \approx-(2/\kappa)\log(1-P) \approx
2.30/\kappa$, or
$\theta_P \approx86.9^\circ/\kappa^{1/2}$.}


Note that, astrophysically, $\kappa$ has a nontrivial interpretation. If
all UHECRs are the same nuclear species (e.g., all protons), then
$\kappa$
depends solely on the magnetic field history experienced by cosmic rays as
they propagate to Earth. If UHECRs are of unknown or mixed chemical
composition, then $\kappa$ conflates magnetic field history and composition.
In a more complicated model, there could be a distribution for the values
of $\kappa$ assigned to UHECRs (accounting for different compositions and
magnetic field histories); the distribution could depend on source direction
(accounting for known magnetic field structure in the Galaxy and
perhaps in
intergalactic space) and on source distance (related to the path length in
intergalactic space).

When estimating $\kappa$ or marginalizing over it, we adopt a log-flat prior
density for $\kappa\in[1,1000]$,
%
%
\begin{equation}
p(\kappa) = \frac{1}{\log1000} \frac{1}{\kappa} \qquad\mbox{for } 1\leq\kappa
\leq1000. \label{k-prior}
\end{equation}
The lower limit corresponds to large angular deflection scales $\sim\!1$ rad,
such as might be experienced by iron nuclei. The upper limit corresponds
to small angular deflection scales $\sim\!1^\circ$, such as might be
experienced by protons with $E\sim100$~EeV.

\subsection{Cosmic ray detection and measurement}
\label{sec:dtxn}

Even though the arrival rate of UHECRs into a unit volume is constant in
time in our model, the expected number per unit time detected from a given
direction will vary as the rotation of the Earth changes the observatory's
projected area toward that direction, as noted above. As a result, the
Poisson intensity function for detectable cosmic rays varies in
time for each source.

Recall that the likelihood function for an inhomogeneous Poisson
point process in time with rate (intensity function) $r(t)$ has the form
%
%
\begin{equation}
\exp(-N_\mathrm{exp}) \prod_i
r(t_i) \delta t, \label{simple-ppp-like}
\end{equation}
where the events are detected at times $t_i$ in detection intervals of size
$\delta t$, and $N_\mathrm{exp}$ is the total expected number in the observing
interval (the integral of the rate over the entire observing interval).
The likelihood function for the cosmic ray data has a similar form, but
with adjustments due to the mark distribution and measurement errors.

If the label and arrival direction for detected cosmic ray $i$ were
known, the
factor in the likelihood function associated with that cosmic ray would be
$F_k \Aperp(\tdrxn_i, t_i) \delta t$, where $k=\lambda_i$.
In reality, both the label and the arrival direction are uncertain; the PAO
analysis pipeline produces a likelihood function for the direction to the
cosmic ray, $\ell_i(\tdrxn_i)$; see equation (\ref{ell-def}).

Introducing the uncertain direction as a nuisance parameter, with a prior
denoted by $\rho_k(\tdrxn_i|\kappa)$, the likelihood factor for
cosmic ray
$i$ when assigned to source $k$ may be calculated by marginalizing; it may
be written as $F_k f_{k,i}\delta t$, with
%
%
\begin{equation}
f_{k,i}(\kappa) = \int \,d\omega_i\, \ell_i (
\omega_i ) \Aperp(\omega_i, t_i)
\rho_k(\omega_i|\kappa). \label{f-def}
\end{equation}
The cosmic ray direction measurement uncertainty is relatively small
($\sim\!1^\circ$) compared to the scale over which the area varies, so we can
approximate $f_{k,i}$ as
%
%
\begin{equation}
f_{k,i}(\kappa) \approx A_i\cos(\theta_i)
\int\ell_i(\omega_i)\rho_k(
\omega_i|\kappa) \,d\omega_i,
\end{equation}
where $\theta_i$ denotes the zenith angle of UHECR $i$ (reported by PAO-10)
and $A_i = A(t_i)$ is the area of the observatory at the arrival time of
UHECR $i$. The integral can be computed analytically,
%
%
\begin{eqnarray}
\label{f-approx}&& \int \,d\omega_i\, \ell_i(
\omega_i) \rho_k(\omega_i|\kappa)
\nonumber
\\[-8pt]
\\[-8pt]
\nonumber
&&\qquad= \cases{
\displaystyle\frac{\kappa_c\kappa}{4\pi\sinh(\kappa_c)\sinh(\kappa)} \frac{\sinh(|\kappa_c n_i+\kappa\hdrxn_k|)}{|\kappa_c n_i+\kappa
\hdrxn_k|}, & \quad $\mbox{if $k\geq1$},$\vspace*{2pt}
\cr
\displaystyle\frac{1}{4\pi}, & \quad $\mbox{if $k=0$}.$}
\end{eqnarray}
The total event rate for cosmic rays with the properties (direction, energy
and arrival time) of detected ray $i$ combines the contributions from each
potential source, that is,
$r(t_i) = \sum_k F_k f_{k,i}(\kappa)$.

To calculate $N_\mathrm{exp}$, we must account
for the observatory's exposure map. The effective exposure given to cosmic
rays from source $k$ throughout the time of the survey depends not just on
the direction to the source, but also on the deflection distribution,
$\rho_k$ (and thus on $\kappa$), since rays from that source will not arrive
precisely from the source direction. The exposure factor for source $k$ is
%
%
\begin{equation}
\varepsilon_k(\kappa) = \int \,d\tdrxn\,\rho_k(\tdrxn|
\kappa) \varepsilon(\tdrxn). \label{eps-def}
\end{equation}
Note that $\varepsilon_k$ has units of area $\times$ time, and for the
isotropic background component ($k=0$), $\varepsilon_0(\kappa)$ is a constant
equal to the sky-averaged exposure (in the notation of
the supplementary material, $\varepsilon_0 = \alpha_T/4\pi$).
To find the total expected number of detected cosmic rays, we sum over
sources: $N_\mathrm{exp} = \sum_{k \ge0} F_k \varepsilon_k(\kappa)$.

The prior probability mass function for the label of a \emph{detected} cosmic
ray is not given by (\ref{label-pmf}); the terms must be weighted
according to
the source exposures. The result is
%
%
\begin{equation}
P(\lambda_i=k|F_0,\Fvec,\kappa) = \frac{F_k\varepsilon_k(\kappa)}{\sum_{j=0}^{N_A} F_j\varepsilon_j(\kappa)}.
\label{label-eps-pmf}
\end{equation}

We now have the ingredients needed to evaluate
equation (\ref{simple-ppp-like}), generalized to include the cosmic ray
marks (directions and labels) and their uncertainties. The resulting
likelihood function is
%
%
\begin{equation}
\like(F_0,\Fvec,\kappa) = \exp \biggl(-\sum
_k F_k\varepsilon_k \biggr) \prod
_i \biggl( \sum_k
f_{k,i} F_k \biggr). \label{like-poisson}
\end{equation}
The product-of-sums factor resembles the likelihood for a finite mixture
model (FMM), if we identify the $f_{k,i}$ factors as the component
densities and the $F_k$ factors as the mixing weights. A common technique
for computing with mixture models is to rewrite the likelihood function
as a
sum-of-products by introducing latent label parameters identifying which
component each datum may be assigned to [see, e.g., \citet{BG88-BayesFMM}].
Following this approach here, the likelihood function can be rewritten
as a
sum over latent assignments of cosmic rays to sources,
%
%
\begin{equation}
\like(F_0,\Fvec,\kappa) = \sum_{\lambda}
\biggl(\prod_k F_k^{m_k(\lambda)}e^{-F_k\varepsilon
_k}
\biggr) \prod_i f_{\lambda_i,i},
\label{like-assoc}
\end{equation}
where $\lambda= \{\lambda_i\}$ and $\sum_\lambda$ denotes an
$N_C$-dimensional sum over all possible assignments of cosmic rays to
sources, and the multiplicity $m_k(\lambda)$ is the number of UHECRs
assigned to source $k$ according to $\lambda$. We suppress the $\kappa$
dependence of $\varepsilon_k(\kappa)$ and $f_{k,i}(\kappa)$ here and elsewhere
to simplify expressions. Note that the $F_k$ dependence (for a given
$\lambda$) is of the same form as a gamma distribution.

Rewriting the previous expression with $(F_T,f)$ in place of $(F_0,F_A)$
and using $F_k = w_kF_A = f w_k F_T$ (for $k\ge1$), we can rewrite
$\like(F_0,\Fvec,\kappa)$ as
%
%
\begin{eqnarray}
\label{eq:like} \like(f,F_T,\kappa) &=& \sum
_\lambda(1-f)^{m_0(\lambda)} f^{N_C-m_0(\lambda)}
F_T^{N_C}
\nonumber
\\[-8pt]
\\[-8pt]
\nonumber
&&\hspace*{14pt}{} \times e^{-F_T [(1-f)\varepsilon_0+f\sum_{k\geq1} w_k\varepsilon
_k  ]} \prod_{k\geq1}
w_k^{m_k(\lambda)} \prod_i
f_{\lambda
_i,i}.
\end{eqnarray}

For computations it will be helpful to have the likelihood function
conditional on the label assignments,
%
%
\begin{equation}
P(D|\lambda,F_0,\Fvec,\kappa) = \exp \biggl(-\sum
_k F_k\varepsilon_k \biggr)
\biggl[ \sum_k F_k
\varepsilon_k \biggr]^{N_C} \prod
_i \frac{f_{\lambda_i,i}}{\varepsilon_{\lambda_i}}, \label{eq:lik-lambda}
\end{equation}
where $k$ runs over the host labels (from 0 to $N_A$), and $i$ runs
over the
UHECR labels (from 1 to $N_C$). We can recover the likelihood for $F_0$,
$\Fvec$ and $\kappa$ by multiplying by the prior for $\lambda$ from
equation (\ref{label-eps-pmf}) and marginalizing, giving
equation (\ref{like-assoc}).

\subsection{\texorpdfstring{Estimating $\kappa$}{Estimating kappa}}

To estimate the deflection parameter, $\kappa$, we need the marginal
likelihood $\like_m(\kappa) = P(D|\kappa) = \int \,dF_T\int \,df
P(D,F_T,f|\kappa)$. The integrand is the product of equation (\ref{eq:like})
and the flux priors. Using the exponential and beta priors described above,
we have that the marginal likelihood for $\kappa$ is
%
%
\begin{eqnarray}
\label{eq:marg} \like_m(\kappa) &=& \sum
_\lambda\frac{\Gamma(N_C+1)\prod_{k\geq1}w_k^{m_k(\lambda)}
\prod_i f_{\lambda_i,i}}{s B(a,b)}
\nonumber
\\[-8pt]
\\[-8pt]
\nonumber
&&\hspace*{14pt}{} \times\int_0^1 \frac{f^{N_C-m_0(\lambda)+a-1}(1-f)^{m_0(\lambda)+b-1}} {
[{1}/{s} + (1-f)\varepsilon_o +
f\sum_{k\geq1}w_k\varepsilon_k ]^{N_C+1}}\,df.
\end{eqnarray}
Computing $\like_m(\kappa)$ requires summing over all possible values of
$\lambda$ which is intractable in practice. In the supplementary material [\citet
{S+13-UHECR-Supp}], we
describe how to use Chib's method [\citet{MR1379473}] to calculate this
marginal likelihood.

\subsection{Model comparison}


To compare rival models, we calculate Bayes factors (ratios of
marginal likelihoods, i.e., posterior odds based on equal prior odds).
Rather than explicitly choosing one model or another
(which would require specification of a loss function), we simply report
Bayes factors as intuitively interpretable summaries of the strength of
evidence in the data for one model over another
[\citet{Kass:Raft:baye:1995}].
This reflects the primarily explanatory (rather than predictive) goals of
astrophysical modeling of UHECR data. With specific predictive goals,
some other model comparison approach could be appropriate (e.g.,
selecting a model via minimizing an information criterion matched to
the predictive goals).

We calculate Bayes factors, both conditioned on $\kappa$ [using marginal
likelihood functions $\like_m(\kappa)$] and after marginalizing over
$\kappa$ [using the log-flat prior of equation (\ref{k-prior}) and
numerical quadrature over $\kappa$].

We consider three models. The null model, $M_0$, assumes that all the
UHECRs come from the isotropic background source population; recall
that it
has no $\kappa$ dependence [see equation (\ref{rho-iso})]. Model $M_1$
allows the UHECRs to come from any of the 17 AGN in the catalog or from the
isotropic background. We also consider another model, $M_2$, in which the
UHECRs may come from the isotropic background or either of the two closest
AGN, Cen A (NGC 5128) and NGC 4945; this model is motivated in part by
recent suggestions that most UHECRs may be heavy nuclei from a single nearby
source, as cited above.
(We also briefly explore a similarly-motivated fourth model that
assigns all
UHECRs to Cen A; as noted below, this model is tenable only for
$\kappa\approx0$.) In order to compare models $M_1$ and $M_2$ (conditioned
on $\kappa$) to the null model, we compute the Bayes factors:
%
%
\begin{equation}
\mathrm{BF}_{10}(\kappa) = \frac{\like_{m,1}(\kappa)}{\like_{m,0}},\qquad \mathrm{
BF}_{20}(\kappa) = \frac{\like_{m,2}(\kappa)}{\like_{m,0}}, \label{BF10+20}
\end{equation}
where $\like_{m,0}$ is the marginal likelihood for the null
model (similar equations hold for models
that marginalize over $\kappa$). The value of $\like_{m,0}$
can be found from equation~(\ref{like-assoc}), noting that for the
null model, there is only one term in the sum over $\lambda$
(with all $\lambda_i = 0$, since the only allowed value of $k$ is $k=0$).
Marginalizing this term over $F_T$ (equal to $F_0$ in this case) gives
%
%
\begin{equation}
\like_{m,0} = \frac{1}{s} \biggl(\frac{s}{s\varepsilon_0+1}
\biggr)^{N_C+1} \Gamma(N_C+1) \times\prod
_i f_{0,i}.
\end{equation}

\subsection{Computational techniques}
\label{subsec:compn}

The principal obstacle to computing with this framework is the combinatorial
explosion in the number of possible associations as the sizes of the
candidate source population and the cosmic ray sample grow. For small
amounts of magnetic deflection, the vast majority of candidate associations
are improbable (they associate well-separated objects with each other). But
there is evidence that UHECRs may be massive (and thus highly charged)
nuclei, which would undergo significant deflection. To probe the full
variety of astrophysically interesting models requires techniques that can
handle both the small- and large-deflection regimes, for catalog sizes
corresponding to current and forthcoming catalogs from PAO.

For parameter estimation within a particular model, we have developed a
Markov chain Monte Carlo (MCMC) algorithm that draws samples of the
parameters $f$, $F_T$ and $\lambda$ from their joint posterior
distribution. The algorithm takes advantage of two features of the models
described above. First, by introducing latent labels, $\lambda$, we could
write the likelihood function in a sum-of-products form,
equation (\ref{like-assoc}), with factors that depend on the fluxes
$F_k$ in the manner of a gamma distribution. Second, the forms of the
likelihood and priors are conjugate for $F_T$ and the labels, so we can find
closed-form expressions for their conditional distributions. These features
enable us to use Gibbs sampling techniques well known in mixture modeling
for sampling the $F_T$ and $\lambda$ parameters. We handle the $f$
parameter using a random walk Metropolis algorithm, so our overall algorithm
is a Metropolis-within-Gibbs algorithm. The supplementary material [\citet
{S+13-UHECR-Supp}] provides details on
its implementation.

We treat the deflection parameter, $\kappa$, specially, considering
a logarith\-mically-spaced grid of values that we condition on. We
did this so that we could explore the~$\kappa$ dependence more
thoroughly than would be possible with posterior sampling of~$\kappa$.
Of course, our Metropolis-within-Gibbs algorithm could be supplemented
with~$\kappa$ proposals to enable sampling of the full posterior.

Finally, using Bayes factors to compare rival models requires computing
marginal likelihoods, which are not direct outputs of MCMC algorithms.
Using a simplified version of our model and modest-sized simulated
data sets, we explored several approaches for marginal likelihood computation
in a regime where we could compute the correct result via direct summation
over all feasible associations. We explored the harmonic mean estimator
(HME), Chib's method and importance sampling algorithms. The HME
performed poorly, often apparently converging to an incorrect result [such
behavior is not unexpected; see \citet{WS12-HMBad}]. Importance sampling
proved inefficient. Chib's method was both accurate and efficient in these
trial calculations, and became our choice for the final implementation.
The supplementary material provides details.

\section{Results}
\label{sec:results}

Recall that the UHECR data reported by PAO-10 are divided into three periods.
The PAO team used an initially larger period 1 sample (including
lower-energy events) to optimize an energy threshold determining which
events to analyze in period 2; the reported period 1 events are only those
with energies above the optimized threshold. The optimization maximized a
measure of anisotropy in the above-threshold period 1 sample. Without access
to the full period 1 sample, we cannot evaluate the impact of this
optimization on our modeling of anisotropy in the reported period 1 data
(nor can we usefully pursue a Bayesian treatment of a GZK energy cutoff
parameter). Because of this complication, we have performed
analyses for various subsets of the data. As our main results, we report
calculations using data from periods 2 and 3 combined (``untuned
data'') and
for periods 1, 2 and 3 combined (``all data''). We also report some
results for each period considered separately, and we use them to
perform a
simple test of consistency of the results across periods, in an effort
to assess the impact of tuning on the suitability of the period 1 data
for straightforward statistical analysis.

\subsection{\texorpdfstring{Results conditioning on the deflection scale, $\kappa$}
{Results conditioning on the deflection scale, kappa}}

We first consider models conditional on the value of the magnetic deflection
scale parameter, $\kappa$, calculating Bayes factors comparing models
and estimates of the association fraction, $f$.

We report model comparison results as curves showing Bayes
factors as functions of $\kappa$. These quantities are astrophysically
interesting but must be interpreted with caution. The actual values of the
conditional Bayes factors can only be interpreted as Bayes factors for a
particular value of $\kappa$ deemed interesting {a priori}. For example,
were one to assume that UHECRs are protons, adopt a particular Galactic
magnetic field model and assume that intergalactic magnetic fields do not
produce significant deflection (which is plausible for protons from local
sources), one would be interested only in large values of $\kappa$ of order
several hundred (corresponding to small angular scales for deflection). On
the other hand, if one presumed that UHECRs are predominantly heavy nuclei,
then deflection by Galactic fields could be very strong, corresponding to
$\kappa$ of order unity (deflection by intergalactic fields might also be
significant in this case). Models hypothesizing that most UHECRs are
heavy nuclei produced by Cen A would fall in this small-$\kappa$ regime.
By presenting results conditional on $\kappa$, various cases
such as these may be considered. Also, the Bayes factor conditioned on
$\kappa$ is proportional to the marginal likelihood for $\kappa$, so the
same curves summarize the information in the data for estimating
$\kappa$ if
it is considered unknown. We plot the curves against a logarithmic
$\kappa$
axis, so they may be interpreted (up to normalization) as posterior
probability density functions based on a log-flat $\kappa$ prior.

\begin{figure}

\includegraphics{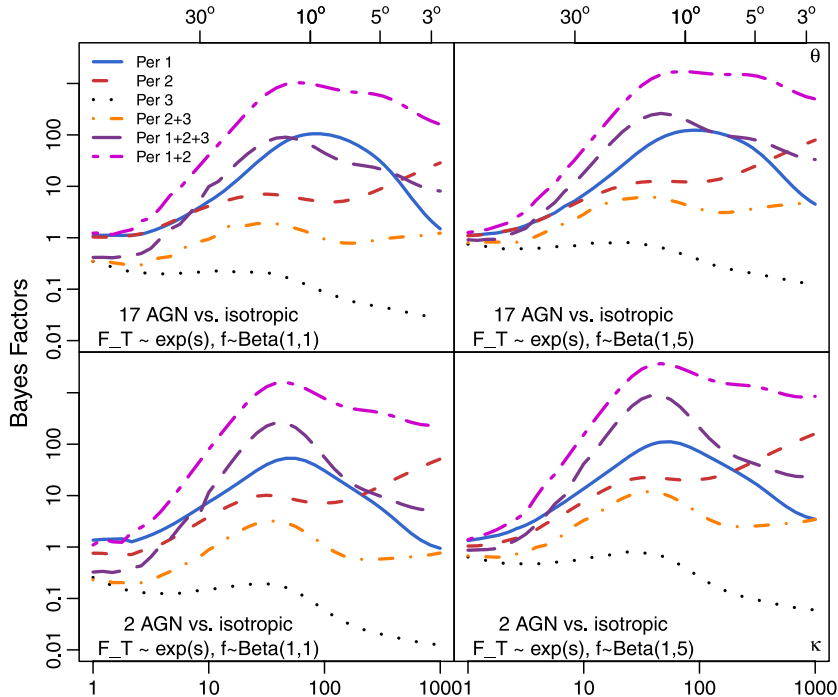}

\caption{Bayes factors comparing the association model with 17 AGN
(top row) or
2 AGN (bottom row) with the null isotropic background model, conditional
on $\kappa$, shown as a function of $\kappa$ (bottom axis) and the
corresponding deflection angle scale, $\theta_P$ (top axis). Results
are shown for various partitions of the data (identified by line style,
identified in the legend) and for two choices of the prior on $f$: a
flat prior (left column) and a Beta$(1,5)$ prior (right column).}
\label{fig:BFplot}
\end{figure}

The Bayes factors comparing models $M_1$ and $M_2$ to $M_0$ for various values
of $\kappa\in[1,1000]$, and for various partitions of the data, are
shown in
Figure~\ref{fig:BFplot}. For cases using only the untuned data
(periods 2, 3
or $2+3$), we find that both $\mathrm{BF}_{10}$ and $\mathrm{BF}_{20}$ [see
equation (\ref{BF10+20})] are close to 1 for all values of $\kappa\in[1,1000]$
for the $\operatorname{Beta}(1,1)$ (uniform) prior for $f$. The Bayes factors are only
a little higher in the case of the $\operatorname{Beta}(1,5)$ prior, indicating the results
are robust to reasonable changes in the $f$ prior. These values imply
that the posterior odds for the association models $M_1$ and $M_2$
versus the null isotropic background model $M_0$ are nearly equal to the
prior odds, indicating the untuned data provide little evidence for or
against either association model versus the isotropic model.

Considering the period 1 data qualitatively changes the results. The
solid (blue) curves in Figure~\ref{fig:BFplot} show the Bayes factor
vs. $\kappa$ results based solely on the period~1 data; there is
strong evidence for association models conditioned on $\kappa$ values
of around 50 to 100.\footnote{A common convention for interpreting
Bayes factors is due to Kass and Raftery, who consider a Bayes factor between
3 and 20 to indicate ``positive'' evidence and between 20 and 150 to indicate
``strong'' evidence [\citet{Kass:Raft:baye:1995}].}
Analyzing the data from all three periods jointly produces the long-dashed
(purple) curves. Using a uniform prior for $f$, we find $\mathrm{BF}_{10}$
attains a
maximum of 90 at $\kappa\approx46$, while $\mathrm{BF}_{20}$ attains a maximum
of 262
at $\kappa\approx38$. Both $\mathrm{BF}_{10}$ and $\mathrm{BF}_{20}$ are larger than 30 for
all $\kappa\in[20,120]$. Both of the association models are strongly
preferred over the null in this range of $\kappa$, while the
comparison is
inconclusive for $\kappa$ outside this range.

The originally published data (in PAO-08) covered periods 1 and 2. For
comparison with studies of that original catalog, Figure~\ref{fig:BFplot}
include curves showing the Bayes factor vs. $\kappa$ based on data
from periods 1 and 2. This partition of the data produces the largest
Bayes factors, $\sim\!1000$ for $\kappa\approx50$. The curves are
qualitatively consistent with accumulation of evidence from periods 1 and~2.\footnote{Note that the Bayes
factor for the $1+2$ partition should not be expected to equal the
product of the Bayes factors based on the periods~1 and~2 partitions,
because the models are composite hypotheses and the data from different
periods generally will favor different values of the model parameters.}
These results amplify what was found in the analysis using all of
the data: the strongest evidence for association comes from the period 1
data. This is troubling because this data was used (along with unreported
lower-energy data) to tune the energy cut defining all of the samples, and
there is no way for independent investigators to account for the
effects of
the tuning on the strength of the evidence in the period 1 data.

\begin{figure}

\includegraphics{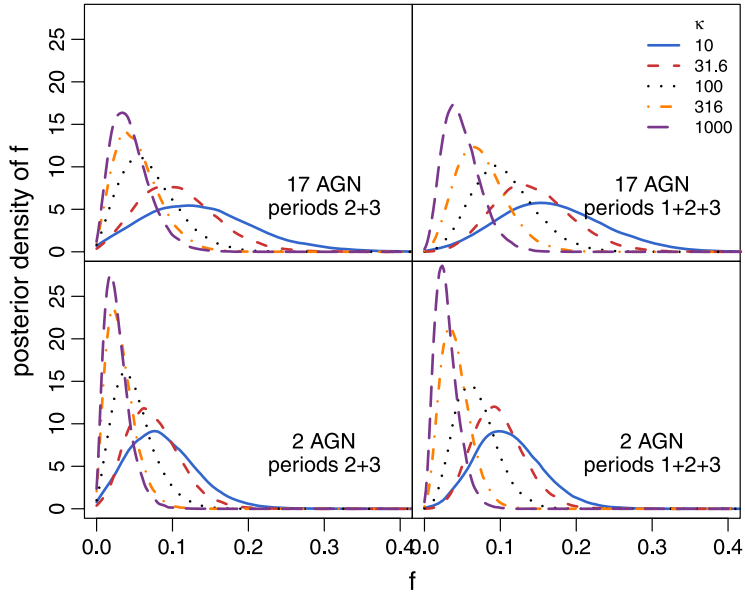}

\caption{Posterior distributions for $f$, conditioned on $\kappa = 10$,
31.6, 100, 316 and 1000.}\label{fig:postf}
\end{figure}

We show marginal posterior densities for $f$ in
Figure~\ref{fig:postf}, for both $M_1$ and $M_2$, using both the untuned
data and using all data. For a given model, the posterior does not
change much when period 1 data are included. The posteriors indicate
evidence for small but nonzero values of $f$, of order a few percent to
20\%. They strongly rule out values of $f>0.3$, indicating that most
UHECRs must be assigned to the isotropic background component in these
models.
This holds even for values of $\kappa$ as small as $\approx\!10$,
corresponding to quite large magnetic deflection scales, as might be
experienced by iron nuclei in typical cosmic magnetic fields. Of
course, when $\kappa=0$ the association models become indistinguishable
from an isotropic background model.

A recent approximate Bayesian analysis [\citet{WMJ11-BayesUHECR}],
based on
a discrete pixelization of the sky, attributed a similar fraction of the
sample of 27 periods~1 and~2 UHECRs to standard candle AGN sources,
considering~$\approx\!900$ AGN within 100 Mpc from the VCV as candidate
sources. But this study adopted an anomalously short GZK-like horizon,
effectively limiting the sample to distances well below 100 Mpc. We compare
our approaches and results in the supplementary material [\citet{S+13-UHECR-Supp}].

\begin{sidewaystable}
\tablewidth=\textwidth
\caption{The posterior probability that each cosmic ray is assigned to each
AGN given $\kappa= 31.62$ and 1000, using cosmic rays from periods $1+2+3$.
Only assignments with probabilities greater than 0.1 are shown. The AGN
identifiers are: 2: NGC 0613; 7: NGC 3621; 11: NGC 4945;
13: NGC 5128 (Cen A); 17: NGC 6300}\label{lambdaTable}
\begin{tabular*}{\textwidth}{@{\extracolsep{\fill}}lcccccccccccccc@{}}
\hline
& \multicolumn{10}{c}{\textbf{17 AGN${} \bolds{+} {}$isotropic}} & \multicolumn{4}{c@{}}{\textbf{2
AGN${} \bolds{+} {}$isotropic}} \\[-6pt]
& \multicolumn{10}{c}{\hrulefill} & \multicolumn{4}{c@{}}{\hrulefill} \\
& \multicolumn{5}{c}{$\bolds{\kappa=31.62}$} & \multicolumn{5}{c}{$\bolds{\kappa=1000}$}& \multicolumn{2}{c}{$\bolds{\kappa=31.62}$} &
\multicolumn{2}{c@{}}{$\bolds{\kappa=1000}$} \\[-6pt]
& \multicolumn{5}{c}{\hrulefill} & \multicolumn{5}{c}{\hrulefill}& \multicolumn{2}{c}{\hrulefill} &
\multicolumn{2}{c@{}}{\hrulefill} \\
\textbf{CR} & \textbf{AGN: 2} & \textbf{7} & \textbf{11} & \textbf{13} & \textbf{17} & \textbf{2} & \textbf{11} & \textbf{13} &
\textbf{16} & \textbf{17} & \textbf{11} & \textbf{13} & \textbf{11} & \textbf{13} \\
\hline
\phantom{0}2 & -- & -- & 0.24 & 0.46 & -- & -- & -- & -- & -- & -- & 0.26 & 0.51 & -- & -- \\
\phantom{0}3 & -- & -- & 0.42 & 0.20 & -- & -- & -- & -- & -- & -- & 0.47 & 0.22 & -- & -- \\
\phantom{0}4 & -- & -- & -- & -- & 0.17 & -- & -- & -- & -- & -- & -- & -- & -- & -- \\
\phantom{0}5 & -- & -- & 0.18 & 0.28 & -- & -- & -- & -- & -- & -- & 0.22 & 0.35 & -- & -- \\
\phantom{0}6 & 0.11 & -- & -- & -- & -- & -- & -- & -- & -- & -- & -- & -- & -- & -- \\
\phantom{0}8 & -- & -- & 0.43 & 0.36 & -- & -- & 0.89 & -- & -- & -- & 0.47 & 0.38 & 0.90
& -- \\
13 & -- & -- & -- & -- & 0.17 & -- & -- & -- & -- & 0.11 & -- & -- & -- & -- \\
14 & -- & -- & 0.47 & 0.27 & -- & -- & -- & -- & -- & -- & 0.51 & 0.29 & -- & --
\\[3pt]
17 & -- & -- & 0.38 & 0.41 & -- & -- & -- & 0.85 & -- & -- & 0.41 & 0.44 & -- &
0.86 \\
18 & -- & -- & -- & 0.15 & -- & -- & -- & -- & -- & -- & -- & 0.20 & -- & -- \\
20 & -- & -- & 0.36 & 0.43 & -- & -- & -- & 0.94 & -- & -- & 0.39 & 0.46 & -- &
0.95 \\
23 & -- & -- & 0.32 & 0.26 & -- & -- & -- & -- & -- & -- & 0.37 & 0.30 & -- & -- \\
26 & -- & 0.17 & 0.10 & 0.19 & -- & -- & -- & -- & -- & -- & 0.15 & 0.27 & -- &
-- \\[3pt]
33 & -- & -- & 0.40 & 0.11 & -- & -- & -- & -- & -- & -- & 0.46 & 0.12 & -- & -- \\
34 & -- & -- & 0.47 & 0.27 & -- & -- & -- & -- & -- & -- & 0.51 & 0.30 & -- & -- \\
36 & -- & -- & 0.21 & 0.35 & -- & -- & -- & -- & 0.48 & -- & 0.24 & 0.42 & -- &
-- \\[3pt]
47 & -- & -- & 0.14 & 0.42 & -- & -- & -- & -- & -- & -- & 0.15 & 0.48 & -- & -- \\
54 & -- & -- & 0.19 & 0.46 & -- & -- & -- & -- & -- & -- & 0.21 & 0.52 & -- & -- \\
55 & 0.15 & -- & -- & -- & -- & 0.34 & -- & -- & -- & -- & -- & -- & -- & -- \\
57 & -- & 0.41 & -- & -- & -- & -- & -- & -- & -- & -- & -- & -- & -- & -- \\
67 & -- & -- & 0.32 & 0.30 & -- & -- & -- & -- & -- & -- & 0.37 & 0.34 & -- & -- \\
\hline
\end{tabular*}
\end{sidewaystable}

The posterior mode is at larger values of $f$ for model $M_1$ (with 17 AGN)
than for $M_2$ (with the two closest AGN), suggesting that there is evidence
that AGN in the G10 catalog besides Cen A and NGC 4945 are sources of
UHECRs. Our multilevel model allows us to address source identification
explicitly, by providing a posterior distribution for possible association
assignments (values of $\lambda$). In Table~\ref{lambdaTable} we show
marginal posterior probabilities for associations that have nonnegligible
probabilities (i.e., $>0.1$), based on models $M_1$ and $M_2$ for two
representative values of $\kappa$ ($\kappa= 31.62$, corresponding to a
$15.5^\circ$ deflection scale, is a favored value for analyses including
period 1 data as shown below; $\kappa=1000$, corresponding to a
$2.7^\circ$
deflection scale, may be appropriate if UHECRs are predominantly protons).
Rows are labeled by cosmic ray number, $i$, and columns by AGN number, $k$;
the tabulated values are $P(\lambda_i=k|\cdots)$. Cosmic rays 17 and
20 (in
period 2) are associated with Cen A (AGN 13) with modest to high probability
in all cases. No other assignments are robust (notably, period 3 has no
robust assignments, despite containing more than three times the number of
cosmic rays as period 2). If UHECRs experience only small deflections, then
besides the two Cen A associations, it is highly probable that cosmic
ray 8
(in period 1) is associated with NGC 4945. For the larger deflection scale,
nearly a quarter of the cosmic rays have candidate associations with
probability $>0.1$, although none of those associations have probability
$>0.5$.
The larger favored value of $f$ for $M_1$ thus reflects the 17 AGN
model, finding enough plausible associations (besides those with Cen A and
NGC 4945) that it is likely that some of them are genuine, even though
it cannot specify which.\looseness=-1

We can also calculate posterior probabilities for multiplet
assignments. In
general, the probability for a multiplet assigning a set of cosmic rays
to a
particular candidate source will not be the product of the
probabilities for
assigning each ray to the source. In Table~\ref{lambdaTable} we see that
CRs 17 and 20 are often commonly assigned to Cen A. As an example, for
$M_1$ with $\kappa=1000$, their separate probabilities for assignment to
Cen A are 0.85 and 0.94, respectively. The probability for a doublet
assignment of both of them to Cen A in this model is 0.80, which
happens to
be nearly equal to the product of their separate (marginal) assignment
probabilities. Were we to marginalize over $\kappa$, the multiplet
probability would differ from the product, since the preferred value
of $\kappa$ differs slightly between these two CRs.


\begin{figure}

\includegraphics{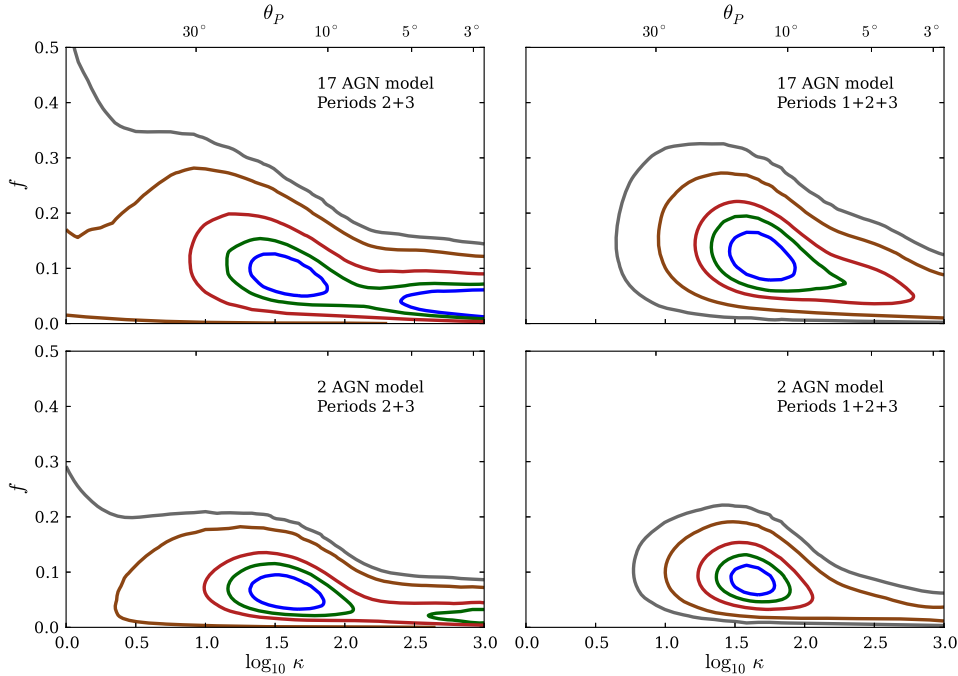}

\caption{Marginal joint posterior distributions for the magnetic deflection
concentration parameter, $\kappa$, and the association fraction, $f$,
considering UHECR data from different periods and candidate host
catalogs of 2
or 17 nearby AGN. Contours bound HPD credible regions of probability
0.25 (blue), 0.5 (green), 0.75 (red), 0.95 (brown) and 0.99 (gray).}
\label{fig:jointkappaf}
\end{figure}

\subsection{\texorpdfstring{Results with $\kappa$ as a free parameter}{Results with kappa as a free parameter}}

Joint marginal posterior distributions for $\log_{10}(\kappa)$ and
$f$ are
shown in Figure~\ref{fig:jointkappaf}, for both association models,
and for
untuned data and all data samples. For the all-data cases, the joint
posterior distribution is unimodal and attains its maximum at ($\kappa
=32$,
$f=0.13$) and ($\kappa=32$, $f=0.09$) for the association model with 17
AGN and 2 AGN, respectively. For untuned data, the joint posteriors are
bimodal with one of the modes at the value of $\kappa$ slightly less
than in
the case of all 3 periods and the other mode at $\kappa\approx1000$,
similar to the plot of Bayes factors in Figure~\ref{fig:BFplot}. The
results from the two samples are more similar than this description may
indicate; they have significant peaks in the same region, but the likelihood
function is relatively flat for the largest and smallest values of
$\kappa$
(this is also apparent in Figure~\ref{fig:BFplot}).

In all cases, the preferred values of $\kappa$ correspond to deflection
scales $\approx\!10^\circ$. As noted above, models of proton
propagation in
cosmic magnetic fields predict deflections of a few degrees.
The posterior distributions for $\kappa$ are comfortably consistent with
such predictions, but they do favor the larger scales that would be
experienced by heavier nuclei. These scales are consistent with the
suggestive evidence from PAO that UHECRs may be comprised of heavier nuclei
than lower-energy cosmic rays.

Values for Bayes factors accounting for $\kappa$ uncertainty are
listed in
Table~\ref{tab:BFtab}, for both association models, and for both individual
and combined data samples (these values are based on the default flat
prior for $f$). We find strong evidence for both association models when
considering all the cosmic ray data. If we exclude the tuned data of period
1, then we see positive evidence for association if we consider only period
2 but positive evidence for the null model if we consider only period
3. If
we pool the untuned data, the data are equivocal. Together, these results
raise concerns about consistency of the data and adequacy of the
models; we
address this further below. These results do not change qualitatively when
we use the alternative prior for $f$ described in Section~\ref{sec3.2}.

\begin{table}
\caption{Overall Bayes factors comparing association models with 17
AGN or
2 AGN to the null isotropic background model, for two different priors for
$f$}\label{tab:BFtab}
\begin{tabular*}{\textwidth}{@{\extracolsep{\fill}}lccccccc@{}}
\hline
& & \multicolumn{6}{c@{}}{\textbf{Data periods used}}\\[-6pt]
& & \multicolumn{6}{c@{}}{\hrulefill}\\
\textbf{Priors for} $\bolds{f}$ & \textbf{Model} & \textbf{1} & \textbf{2} & \textbf{3} & \textbf{1\&2} & \textbf{2\&3} & \multicolumn{1}{c@{}}{\textbf{1\&2\&3}}\\
\hline
$\operatorname{Beta}(1,1)$ & 17 AGN & 31 & 6.5 & $0.15 = 1/6.7$ & \phantom{0}370 & 0.99 & \phantom{0}26\\
& 2 AGN & 15 & 9.9 & $0.11 = 1/9.1$ & \phantom{0}440 & 1.1\phantom{0} & \phantom{0}51\\[3pt]
$\operatorname{Beta}(1,5)$ & 17 AGN & 39 & 15\phantom{00.} & $0.52 = 1/1.9$ & \phantom{0}710 & 3.4\phantom{0} & \phantom{0}79\\
& 2 AGN & 32 & 28\phantom{00.} & $0.42 = 1/2.4$ & 1100 & 4.1\phantom{0} & 180\\
\hline
\end{tabular*}
\end{table}

\begin{figure}[b]

\includegraphics{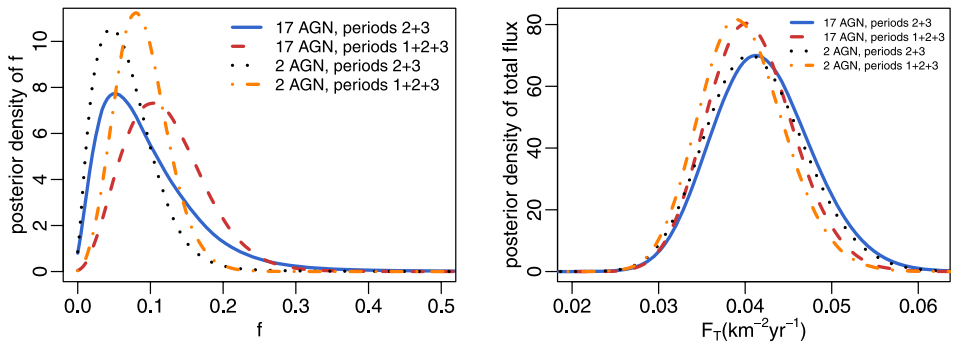}

\caption{Marginal posterior distributions for $f$ (the fraction of
UHECRs associated with AGN in candidate catalogs) and $F_T$ (the total
flux), considering UHECR data from different periods and models
associating UHECRs with either 2 or 17 nearby AGN.}
\label{fig:posterior}
\end{figure}


Marginal posterior distributions for $f$ and for $F_T$ are shown in
Figure~\ref{fig:posterior}. For the untuned data, the posterior mode of
$f$ is 0.051 for $M_1$ (17 AGN) and 0.047 for $M_2$ (2 AGN);
the 95\% highest density credible intervals for $f$ are $[0, 0.23]$
and $[0.002, 0.145]$, respectively. Using all of the data, the
distributions shift to somewhat larger values of $f$; the
posterior mode of $f$ is 0.11 for $M_1$ and 0.08 for $M_2$, and $f=0$
has a significantly smaller density. However, the uncertainties are
large enough that the $f$ estimates are consistent with each other. The
posterior distributions for $F_T$ are very similar in all models. The
peaks are a little higher and the widths of the peaks are smaller when
we consider the cosmic rays from periods~1--3, as expected, since we
have more data. The posterior modes correspond to total fluxes of about
0.04 km$^{-2}$ yr$^{-1}$ in all cases.

%
%
%
%

\subsection{Single-source models}

Some investigators have suggested
that UHECRs are all heavy nuclei from a single source---the nearest AGN,
Cen A---with the apparent approximate isotropy of arrival directions
a consequence of strong deflection [\citet{B+09-CenA,GBdS10-CenA,BdS12-CenA}].
This hypothesis is motivated by the ability to fit the all-sky energy
spectrum above 50~EeV with models that predict negligible proton content.
The marginal posterior distributions for $f$ in
Figure~\ref{fig:postf} strongly rule out values of $f>0.3$
even for large magnetic deflection scales; such models are too
anisotropic. These results are for models
allowing multiple sources, but they suggest
that a model assigning \emph{all} UHECRs to a
single nearby source may be untenable for astrophysically
plausible deflection scales.
In the supplementary material [\citet{S+13-UHECR-Supp}] we briefly explore models
attributing all UHECRs to
Cen A, as a function of $\kappa$. The $\kappa=0$ case corresponds to
a truly
isotropic distribution for UHECR directions, and thus has a Bayes factor
(vs. the background model) of unity. We show that the Bayes factor
decreases quickly as $\kappa$ grows; even small amounts of anisotropy
toward Cen A are contraindicated by the data. Models with $\kappa
\gtrsim0.5$,
that is, with deflection angular scales $<90^\circ$, are strongly ruled
out. Larger deflection scales require Galactic field strengths that are
surprisingly large [see equations (\ref{dflxn-reg}) and (\ref{dflxn-turb})].
These results indicate that Cen A single-source models are ruled out
unless very large deflection scales can be justified, and even then they
are disfavored. More details are in the supplementary material.

\subsection{Model checking}

In the supplementary material [\citeauthor{S+13-UHECR-Supp}\break (\citeyear{S+13-UHECR-Supp})] we describe results of
two types of
tests of our models, motivated by period-to-period variability of some of
the results reported above.

First, we performed simple change point analyses to see whether the
period-to-period variation of the Bayes factors for association vs. isotropy
indicates the population-level properties of the detected cosmic rays vary
from period to period. We compared versions of $M_1$ and $M_2$ that allow
model parameters to change between periods to versions that keep the
parameters the same for all periods. We find that there is no significant
evidence for variability of model parameters from period to period.


Second, we performed predictive checks to see whether the period-to-period
Bayes factor variations are surprising in the context of either the
null or
association models, essentially using the Bayes factors as goodness-of-fit
test statistics.
We simulated data from the null (isotropic) model and compared the Bayes
factors based on the observed data with those found in the simulations; we
did the same for a representative association model.
We find that Bayes factors favoring association as large as that
found with the period 1 PAO data are unlikely for
isotropic models.
This implies the distribution of directions in the period 1 sample is
anisotropic, but the calculation does not address whether this may be
due to
tuning or to genuine anisotropy.
For association models, the large Bayes factors for periods 1 and 2,
and the small Bayes factor in period 3, are not individually surprising.
But it is very surprising to see a combination of large Bayes factors
for each of the two small subsamples, and a small Bayes factor for the large
subsample. The full data set thus is not comfortably fit by either
isotropic or association models. We discuss this further below.

The simulations used for model checking, with known ``ground truth,'' also
provide some insight into the frequentist calibration of inferences,
for example, the
coverage of credible regions, and the accuracy of CR--AGN associations as
a function of the association probability. This is discussed further
in the supplementary material.

\section{Summary and discussion}
\label{sec:summary}

We have described a new multilevel Bayesian framework for modeling the
arrival times, directions and energies of UHECRs, including statistical
assessment of directional coincidences with candidate sources.
Our framework explicitly models cosmic ray emission, propagation (including
deflection of trajectories by cosmic magnetic fields) and detection. This
approach cleanly distinguishes astrophysical and experimental processes
underlying the data. It handles uncertain parameters in these processes via
marginalization, which accounts for uncertainties while allowing use of all
of the data (in contrast to hypothesis testing approaches that optimize over
parameters, requiring holding out a subset of the data for tuning).
We demonstrated the framework by implementing calculations with simple but
astrophysically interesting models for the 69 UHECRs with energies above
55~EeV detected by PAO and reported in PAO-10. Here we first summarize
our findings based on these models, and then describe directions for
future work.


\subsection{Astrophysical results}

We modeled UHECRs as coming from either nearby AGN (in a volume-limited
sample including all 17 AGN within 15 Mpc) or an isotropic background
population of sources; AGN are considered to be standard candles in
our models. We thoroughly explored three models. In $M_0$ all CRs come
from the isotropic background; in $M_1$ all CRs come from either a
background or one of the 17 closest AGN; in $M_2$ all CRs come from either
a background source or one of the two closest AGN (Cen A and NGC 5128,
neighboring AGN at a distance of 5 Mpc). The data were reported in three
periods. Data from period 1 were used to tune the energy threshold defining
the published samples in all periods by maximizing an index of
anisotropy in
period 1. Out of concern that this tuning compromises the data in period 1
for our analysis, we analyzed the full data set and various subsamples,
including an ``untuned'' sample omitting period 1 data.

Using \emph{all} of the data, Bayes factors indicate there is strong evidence
favoring either $M_1$ or $M_2$ against $M_0$ but do not discriminate between
$M_1$ and $M_2$. The most probable models associate about 5\% to 15\% of
UHECRs with nearby AGN and strongly rule out associating more than
$\approx\!25$\% of UHECRs with nearby AGN. Most of the high-probability
associations in the 17 AGN model are with the two closest AGN.

However, if we use only the \emph{untuned} data, the Bayes factors are
equivocal (although the most probable association models resemble those
found using all data). If we subdivide the untuned data, we find positive
evidence for association using the period 2 sample, but weak evidence
\emph{against} association using the much larger period 3 sample.
Together, these
results suggest that the statistical character of the data may differ from
period to period, due to tuning of the period 1 data or other causes.

One way to explore this is to ask whether the data from the various periods
are better explained using models with differing parameter values rather
than a shared set of values. We investigated this via a change-point
analysis that considered the time points bounding the periods as candidate
change points. The results are consistent with the hypothesis that the
parameters do \emph{not} vary between periods, justifying using the combined
data for these models. This suggests the variation of the Bayes factors
across periods is a consequence of the modest sample sizes. However, the
change-point analysis does not address the possibility that none of the
models is adequate, with model misspecification being the cause of the
apparently discrepant Bayes factors.

We used simulated data from both the isotropic model and
high-probability association models to perform predictive checks of our
models, using the Bayes factors based on subsets of the data as test
statistics. Simulations based on the isotropic model indicate that large
Bayes factors favoring association are unlikely for \emph{untuned}
samples of
the size of the period 1 sample. Simulations based on representative
association models indicate that such Bayes factors are not surprising
for samples of the size of period 1, considered in isolation. But
the observed pattern of large Bayes factors for the subsamples in
periods 1 and 2, and a small Bayes factor for the much larger period 3
subsample, is very surprising. The full data set thus is not fit
comfortably by either isotropic models or standard candle association
models.
Whether the effects of tuning could explain the apparent inconsistencies
remains an open question that is not easy to address without access to the
untuned data.

Restricting to the untuned data (periods 2 and 3), the pattern of Bayes
factors is consistent with both isotropic models and representative standard
candle association models.
The best-fitting association models assign a few percent of UHECRs to nearby
AGN; at most $\approx\!20$\% may be associated with AGN, with the remainder
assigned to sources drawn from an isotropic distribution.
Magnetic deflection angular scales of $\approx\!3^\circ$
to $30^\circ$ are favored.
Models that assign a large fraction of UHECRs to a single nearby source (e.g.,
Cen A) are ruled out unless very large deflection scales are specified a
priori, and even then they are disfavored.

Even restricting to results based on the untuned data, we hesitate to offer
these models as astrophysically plausible explanations of the PAO UHECR
data, both because of how important the problematic period 1 sample is in
the analysis and because of astrophysical limitations of the models
considered here and elsewhere. In particular, the high-probability models
assign the vast majority of UHECRs to sources in an isotropic distribution.
But the observation by PAO of a GZK-like cutoff in the energy spectrum of
UHECRs suggests that UHECRs originate from within $\sim\!100$ Mpc,
where the distribution of both visible matter (galaxies) and dark
matter is
significantly \emph{an}isotropic. If most or all UHECRs are protons,
so that
magnetic deflection is not very strong, an isotropic distribution of UHECR
arrival directions is implausible. It then may be the case that some of the
strength of the evidence for association with nearby AGN is due to the
``straw man'' nature of the isotropic alternative. On the other hand, if
most UHECRs are heavy nuclei, then strong magnetic deflection could
isotropize the arrival directions. The highest probability association
models have relatively small angular deflection scales, but it could be that
the few UHECRs that these models associate with the nearest AGN happen
to be
protons or very light nuclei. Future models could account for this by
allowing a mixture of $\kappa$ values among cosmic rays, as noted in
Section~\ref{sec:dflxn}.
In addition, the standard candle cosmic ray intensity model adopted here
and in other studies very likely artificially constrains inferences.


\subsection{Future directions}

All of these considerations indicate a more thorough exploration of
UHECR production and propagation models is needed. We thus consider
the analyses here to be a demonstration of the utility and feasibility
of analyzing such models within a multilevel Bayesian framework, and
not a definitive astrophysical analysis of the data.
We are pursuing more complex models separately, expanding on the present
analysis in four directions.

First, we are considering larger, statistically well-characterized catalogs
of potential hosts, for example, the recently-compiled catalog of X-ray selected
AGN detected by the Burst and Transient (BAT) instrument on the \emph{Swift}
satellite, a catalog considered by PAO-10.

Second, we are building more realistic background distributions, for example,
by using the locations of nearby galaxy clusters or the entire nearby
galaxy distribution, to build smooth background densities (e.g., via kernel
density estimation, or fitting of mixture or multipole models).

Third, we are considering richer luminosity function models, including models
assigning a distribution of cosmic ray intensities to all candidate sources
and models that place some sources in ``on''
states and the others ``off.'' The latter models are motivated both by the
possibility of beaming of cosmic rays and by evidence for AGN intermittency
in jet substructure, and could enable assignment of significant numbers of
UHECRs to both distant and nearby sources.

Finally, more complicated deflection models are possible.
For example, we have developed a class of ``radiant'' models that produce
correlated deflections (as seen in some astrophysical simulations). For a
radiant model, each source has a single guide direction associated with it,
drawn from a Fisher distribution centered at the source direction, with
concentration $\kappa_g$; the guide direction serves as a proxy for
the shared
magnetic deflection history of cosmic rays from that source. Each
cosmic ray
associated with that source then has its arrival direction drawn from an
independent Fisher distribution centered about the guide direction, with
concentration potentially depending on cosmic ray energy and source distance;
this distribution describes the effect of the deflection history unique
to a
particular cosmic ray. The resulting directions for a multiplet will cluster
along a ray pointing toward the source. The resulting joint
distribution for
the directions in a multiplet (with the guide direction marginalized) is
exchangeable but not independent.

For the current, modest-sized UHECR catalog, the complexity of
some of these generalizations is probably not warranted. But PAO is
expected to operate for many years, and the sample is continually
growing in
size. Making the most of existing and future data will require not only
more realistic models, but also more complete disclosure of the data.
In particular, a fully Bayesian treatment---including modeling of the energy
dependence in the UHECR flux and deflection scale---requires data
uncorrupted by tuning cuts. Further, the most accurate analysis should use
event-specific direction and energy uncertainties (likelihood summaries),
rather than the typical error scales currently reported. We hope
our framework helps motivate more complete releases of future PAO data.


\section*{Acknowledgment}
We are grateful to Paul Sommers for helpful conversations about the
PAO instrumentation and data reduction and analysis processes.

\begin{supplement}[id=suppA]
\stitle{Technical appendices}
\slink[doi]{10.1214/13-AOAS654SUPP} 
\sdatatype{.pdf}
\sfilename{aoas654\_supp.pdf}
\sdescription{The online supplement contains six technical appendices
with detailed material on the following topics:
\begin{longlist}[F.]
\item[A.] Auger observatory exposure;
\item[B.] Propagation effects on cosmic ray energies;
\item[C.] Algorithm for Markov chain Monte Carlo;
\item[D.] Cen A single-source model;
\item[E.] Comparison with prior Bayesian work;
\item[F.] Model checking.
\end{longlist}}
\end{supplement}



%

\printaddresses

\end{document}